\documentclass{aa}

\usepackage{graphicx}
\usepackage{txfonts}
\def\kms {\hbox{${\rm km\,s}^{-1}$}}

\def\as {\hbox{$^{\prime\prime}$}}

\def\ffas {\hbox{$\,.\!\!^{\prime\prime}$}}

\def \ga{\mathrel{\mathchoice   {\vcenter{\offinterlineskip\halign{\hfil
$\displaystyle##$\hfil\cr>\cr\sim\cr}}}
{\vcenter{\offinterlineskip\halign{\hfil$\textstyle##$\hfil\cr
>\cr\sim\cr}}}
{\vcenter{\offinterlineskip\halign{\hfil$\scriptstyle##$\hfil\cr
>\cr\sim\cr}}}
{\vcenter{\offinterlineskip\halign{\hfil$\scriptscriptstyle##$\hfil\cr
>\cr\sim\cr}}}}}

\begin{document}

   \title{ NH$_{3}$ $(1,1)$ hyperfine intensity anomalies in infall sources }

   \titlerunning{NH$_{3}$ $(1,1)$ hyperfine intensity anomalies in infall sources}

   \author{Gang Wu
          \inst{1,2}
          \and
          Christian Henkel
          \inst{2,1}
          \and
          Dongdong Zhou
          \inst{1}
          \and
          Friedrich Wyrowski
          \inst{2}
          \and
          Karl M. Menten
          \inst{2}
          \and
          Jarken Esimbek
          \inst{1}
           }

   \institute{Xinjiang Astronomical Observatory, CAS  150, Science 1-Street Urumqi, Xinjiang 830011, China\\
   \email{wug@xao.ac.cn}
   \and
    Max-Planck-Institut f\"{u}r Radioastronomie, Auf dem H\"{u}gel 69, 53121, Bonn, Germany}

   \date{Received ...; accepted ...}

% \abstract{}{}{}{}{}
% 5 {} token are mandatory

  \abstract
   {Identifying infall motions is crucial for our understanding of accretion processes in regions of star formation. The NH$_{3}$ $(1,1)$ hyperfine intensity anomaly (HIA) has been proposed to be a readily usable tracer for such infall motions  in star-forming regions harboring young stellar objects at very early evolutionary stages.
   In this paper, we seek to study the HIA toward fifteen infall candidate regions to assess its reliability as an infall tracer. By using deep observations of the NH$_{3}$ $(1,1)$ transition with the Effelsberg 100\,m telescope, HIAs have been identified toward all the targets.% are determined.
   Fourteen out of fifteen sources exhibit anomalous intensities either in the inner or outer satellite lines.
   All the derived HIAs conform to the framework of the existing two models, namely, hyperfine selective trapping (HST) and  systematic contraction or expansion motion (CE) models.
   In our sample of infall candidates, a majority of the HIAs remain consistent with the HST model.
   Only in three targets, the HIAs are consistent with infall motions under the CE model.
   Thus HIAs could be used as an infall tracer but seem not highly sensitive to infall motions in our single-dish data.
   Nevertheless, the emission could be blended with emission from outflow activities.
   HIAs consistent with the HST model show stronger anomalies with increasing kinetic temperatures ($T_{\rm K}$), which is expected by the HST model.
   On the other hand, HIAs consistent with infall motions show little dependence on %seem to show relatively constant values against
   $T_{\rm K}$.
   Therefore, HIAs may preferably trace %to serve as
   infall of cold gas.}

   \keywords{ISM: clouds -- ISM: kinematics and dynamics -- ISM: molecules -- stars: formation}

   \maketitle
%
%-------------------------------------------------------------------

\section{Introduction}
\label{intro}

Accretion is a fundamental phenomenon in the process of star formation \citep[e.g.][]{1977ApJ...214..488S, 2004RvMP...76..125M}.
However, observational evidence of gas infall motions has remained inconclusive \citep[e.g.][]{1999ARA&A..37..311E, 2015ApJ...814...22E}.
Blue-skewed line profiles of optically thick lines serve as readily accessible tracers for infall motions, relying on excitation gradients within the clump \citep[e.g.][]{1977ApJ...214L..73L, 1993ApJ...404..232Z}.
Given higher excitation temperatures toward the center, the inward motion of the clump will manifest itself in more pronounced blueshifted emission \citep[][]{1993ApJ...404..232Z, 1999ARA&A..37..311E}.
Nevertheless, there are also alternative explanations, such as chemical variations and rotation \citep[e.g.][]{2003cdsf.conf..157E}.
At the same time, blue-skewed line profiles observed by single-dish telescopes may be blended with red-skewed profiles caused by outflow activities.
Additionally, the requirement of elevated excitation toward the center implies that the blue-skewed emission as an infall tracer becomes less pronounced for objects in very early evolutionary stages with lower radial changes of the excitation temperature.

Searching for redshifted absorption features against radio or millimeter continuum emission appears to be a promising approach \citep[e.g.][]{2001ApJ...562..770D,2012A&A...542L..15W,2022A&A...658A.192Y}.
The inverse P-Cygni profile, characterized by redshifted absorption and blueshifted emission, is a relatively robust infall indicator, because these profiles largely confirm that the gas moving inwards (redshifted) is in front of the central source.
Observations with high angular resolution interferometers are usually required to confirm infall toward  the central continuum source \citep[e.g.][]{2022A&A...659A..81B}.
Because of the optically thick nature of the dust emission at frequencies $\ga$1\,THz, absorption lines are more readily observed in this frequency range.
\citet{2016A&A...585A.149W} employed the Stratospheric Observatory for Infrared Astronomy (SOFIA) at $\sim$1.8\,THz and detected redshifted absorption features toward high-mass star formation regions.
Additionally, they noted that these redshifted absorption features are often not significantly contaminated by redshifted outflowing gas. % at larger wavelengths.
However, acquiring new data for  these high-frequency absorption lines, which are beyond the capabilities of ground-based telescopes, is not possible after the end of the SOFIA mission.

In all aforementioned methods, optically thin lines are usually required to trace the systemic velocities, delineating the boundary between the blue- and redshifted features and ruling out the possibility of two separate velocity components.
Continuous efforts have been made to improve our capacity for detecting infall motions, aiming to achieve a comprehensive understanding of these motions in star-forming regions across a wider range of evolutionary stages and spatial scales.
In this study, we test the viability of using ammonia hyperfine intensity anomalies (HIAs) as a promising infall tracer.

The microwave frequency inversion transitions of
ammonia (NH$_{3}$) are frequently used tracers in studies of star-forming regions, molecular clouds, and nearby galaxies \citep[e.g.,][]{1998ApJ...505L.151Z, 2005A&A...440..893H, 2014ApJ...790...84L,2018A&A...616A.111W}.
The ground-state para NH$_{3}$ $(J, K) = (1,1)$ transition comprises five distinct groups of hyperfine structure (hfs) components, originating from electric quadrupole splitting. These components include the main line ($\Delta F_{1}$=0, where $F_{1}$ is the quantum number of the electric quadrupole coupling hyperfines) and four satellite lines ($\Delta F_{1}$=$\pm$1), with two on each side of the main line \citep{1983ARA&A..21..239H}.
The presence of weaker magnetic spin-spin interactions results in a total of 18 hyperfine components distributed over the profiles of the five lines \citep[e.g.][]{1977ApJ...215L..35R}.
Under conditions of local thermodynamic equilibrium (LTE) and optically thin emission, the two inner satellite line groups (ISLs) and outer satellite line groups (OSLs) are anticipated  to have the same intensities (26\% for each ISL and 22\% for each OSL of the main line intensity) \citep[e.g.,][]{1983ARA&A..21..239H}.
However, the expectation of equal intensities for the ISLs and OSLs is not always valid \citep[e.g.][]{1977ApJ...214L..67M, 2015ApJ...806...74C}.
Non-LTE populations between the sub-states of the NH$_{3}$ $(1,1)$ level, which give rise to HIAs, can be attained by (1) hyperfine selective trapping (HST) from NH$_{3}$ $(2,1)$ to $(1,1)$ levels \citep[e.g.,][]{1977ApJ...214L..67M,1985A&A...144...13S} and/or (2) systematic contraction or expansion motions (CE) \citep[e.g.,][]{2001A&A...376..348P}.
It is possible to distinhguish  between these two models, because they predict different relative intensities of the satellite lines (see Section \ref{hia-infall}).
Under the CE model, redshifted/blueshifted photons (due to systematic motions) emitted from one hyperfine transition can be absorbed by another one, resulting in substantial changes in the level populations of the NH$_{3}$ $(1,1)$ sub-levels.
Contraction/expansion can enhance the emission of the two satellite lines on the blue/red side, while suppressing those on the other side \citep[e.g.,][]{2001A&A...376..348P}.
Thus, the HIA is expected to serve as a tracer of systematic motions without relying on detailed analysis of line shapes \citep[][]{2001A&A...376..348P}.

To summarize, the HIA, as an infall tracer, exhibits two enhanced blueshifted satellite lines, resembling a discrete version of the blue-skewed profile, albeit with different underlying physics.
The presence of five well-separated components in the NH$_{3}$ $(1,1)$ transition enables the straightforward identification of enhanced emission.
Furthermore, the HIA is based on the cross-absorption between the hyperfine transitions. In principle, higher excitation toward the center is not required.
For example, in the simulations conducted by \citet{2001A&A...376..348P}, obvious HIAs were reproduced in an infalling core with a constant temperature of 15\,K.
This makes the HIA a promising infall tracer of star-forming regions at very early evolutionary stages and also of large scale accretion.
In contrast,  blue-skewed profiles can reliably trace the infall motion around a hot core  \citep[][]{2003cdsf.conf..157E}.

Nonetheless, the relative contribution of systematic motions to HIAs remains a matter of debate \citep[e.g.][]{2007MNRAS.379..535L,2018A&A...609A.125W,2020A&A...640A.114Z}.
In these studies, HIAs tend to be consistent with the HST model.
However, infall motions of the targets in these studies are not confirmed.
Hence, in this paper, we seek to investigate HIAs toward fifteen infall candidates, indicated by blue-skewed or redshifted absorption profiles, to test whether the HIA can be used as an infall tracer.

\section{Targets and observations}
\label{obs}

\subsection{Targets}

We selected 15 infall candidates from the literature (see Table \ref{tab:obs}).
Blue-skewed profiles were extensively observed in B335, such as by the Institute de Radioastronomie Millim\'{e}trique (IRAM) 30\,m telescope in H$_{2}$CO (2$_{12}$--1$_{11}$), CS (2--1), (3--2), (5--4) and by the Caltech Submillimeter Observatory (CSO) 10.4 m telescope in HCO$^{+}$ (3--2) with resolutions ranging from 11$\as$  to 28$\as$ \citep[e.g.][]{1993ApJ...404..232Z}.
Inverse P-Cygni profiles were also seen in Atacama Large Millimeter/submillimeter Array (ALMA) HCN (4--3) and HCO$^{+}$ (4--3) spectra with a resolution of about 0$\ffas$5 \citep{2015ApJ...814...22E}.
However, recent high-resolution ALMA observations  of optically thin tracers revealed the presence of two velocity components, which could contribute to the double-peaked line profiles \citep[e.g.][]{2021A&A...653A.166C}.

Infall indicators were also widely identified in G031.41+0.31. For example, a redshifted absorption profile was identified in the SOFIA NH$_{3}$ 3$_{2+}$--2$_{2-}$ line with a resolution of about 16$\as$ \citep{2012A&A...542L..15W}.
Meanwhile, blue-skewed profiles were detected in the  HCO$^{+}$ (4--3),  HNC (4--3), HCN (4--3) lines
observed with the using the Atacama Pathfinder Experiment 12 meter submillimeter telescope (APEX) and the CS (2--1) line observed with the IRAM 30 m telescope  with resolutions ranging from 17$\as$  to 27$\as$ \citep{2016A&A...585A.149W}.
Inverse P-Cygni profiles were also detected in  dense cores with ALMA spectra of CH$_{3}$CN (12--11) and H$_{2}$CO (3$_{0,3}$--2$_{0,2}$), (3$_{2,2}$--2$_{2,1}$), and (3$_{2,1}$--2$_{2,0}$) with resolutions of about 0$\ffas$1 \citep[e.g.][]{2022A&A...659A..81B}.

The inverse P-Cygni profile in IRAS\,18360-0537 was observed near the dust peak MM1 using the Submillimeter Array (SMA) in the CN (2--1) line with a resolution of about 1$\ffas$4 \citep{2012ApJ...756..170Q}.
Blue-skewed profiles were also identified with the CSO in HCN (3--2) and  the Arizona Radio Observatory (ARO) 12\,m telescope in HCO$^{+}$ (1--0) with resolutions of 27$\as$ and 70$\as$, respectively  \citep{2018ApJS..235...31Y}.

G023.21$-$0.3, G034.26+0.2, and G035.20$-$0.7 were selected from \citet{2016A&A...585A.149W}. In these three targets, redshifted absorption features were observed using the SOFIA in the NH$_{3}$ 3$_{2+}$--2$_{2-}$  line at a resolution of about 16$\as$. Furthermore, blue-skewed profiles were identified in all three targets through in HCO$^{+}$ (4-3) with APEX with a resolution about 17$\as$ \citep{2016A&A...585A.149W}.
As also reported in \citet{2016A&A...585A.149W}, in the case of G023.21$-$0.3, APEX HNC (4-3) observations with a resolution of 16$\as$ revealed blue-skewed profiles. Similarly, in the case of G034.26+0.2, APEX HCN (4-3) observations with a resolution of 17$\as$ showed blue-skewed profiles. As for G035.20$-$0.7, IRAM 30m HCO$^{+}$ (1--0) observations with a resolution of 28$\as$ revealed the presence of blue-skewed profiles.

BGPS\,3604, BGPS\,4029, and BGPS\,5021 were selected from \citet{2018ApJ...862...63C}. Blue-skewed profiles were observed in the three targets using ARO 12\,m  single-pointing HCO$^{+}$ (1--0) observations, with a resolution of about 68$\as$ \citep{2018ApJ...862...63C}.
\citet{2021ApJ...922..144Y} mapped G029.60$-$0.63, G053.13+0.09, G081.72+0.57, G082.21$-$1.53, G121.31+0.64, and G193.01+0.14 with the IRAM 30m telescope at a resolution of about 28$\as$. In all of  these six targets, blue-skewed profiles were detected in the HCO$^{+}$ (1--0) line toward the positions showing strongest H$^{13}$CO$^{+}$ (1--0) emission \citep{2021ApJ...922..144Y}.

\subsection{NH$_{3}$ observations}

Deep NH$_{3}$ observations (PI: Gang Wu, project ID: 15-21) were conducted in February 2022 with the Effelsberg 100\,m telescope\footnote{The 100-m telescope at Effelsberg is operated by the Max-Planck-Instiitut f{\"u}r Radioastronomie (MPIfR) on behalf of the Max-Planck-Gesellschaft (MPG).}.
The K-band double-beam and dual-polarization receiver was employed as the frontend.
The facility Fast Fourier Transform Spectrometer (FFTS) was used as backend, which offered two frequency windows with bandwidths of 300\,MHz and 65536 channels each.
This results in a channel width of about 4.6\,kHz, corresponding to a velocity spacing of about 0.06\,$\kms$ at 23.7\,GHz.
The three metastable NH$_{3}$ $(J,K) = (1,1), (2,2),$ and $(3,3)$ lines, along with the $J$$_{K_{a}}$$_{K_{c}}$\,=6$_{16}$-5$_{25}$ water maser transition, were observed simultaneously.
Spectral calibration was applied, following the method described  by \citet{2012A&A...540A.140W}.
NGC\,7027 was used to obtain the initial pointing and focus corrections and to calibrate the spectral line flux, assuming a continuum flux density of 5.7\,Jy at 23.7 GHz \citep{1994A&A...284..331O}.
At the NH$_3$ frequencies, the full width at half maximum (FWHM) beam size is about 37$\as$ and the main beam efficiency is 60\%.
The conversion factor from the flux density scale to the main beam brightness temperature is 1.73\,K\,Jy$^{-1}$.
The focus was checked every few hours and pointing was calibrated every hour by observing  nearby compact continuum sources.
A position-switching mode was used in the observations with the off position at an offset of 900$\as$ of each target in azimuth.
Since achieving good signal-to-noise ratios for the NH$_{3}$ $(1,1)$ satellite lines is crucial to establish a robust HIA \citep[e.g.][]{2020A&A...640A.114Z},
the total on\,+\,off  integration time on each target exceeds 70 minutes and 1-$\sigma$ noise levels are about 50\,mK on a $T_{\rm mb}$ scale for a channel width of 0.06 $\kms$ (see Table \ref{tab:obs}).
Note that NH$_{3}$ $(1,1)$ hyperfine features were all measured simultaneously, thus ensuring an accurate relative calibration.

The GILDAS/CLASS software developed by IRAM was mainly used for the data reduction.
All of the spectra have flat and gently varying baselines so that only linear baselines had to be subtracted.

   \begin{table*}
    \caption{Targets and observational parameters.}
    \label{tab:obs}
    \centering
    \begin{tabular}{cccccccccc}     % 7 columns
    \hline\hline
    No. & Target & R.A. (J2000) &  DEC (J2000)  &  D(kpc)  &   on-of  (min)  &  $\sigma_{\rm BL}$ (K)\tablefootmark{$\dag$} & $T_{\rm K}$ & $\tau_{\rm ISL}$\tablefootmark{$\dag\dag$}  & $\tau_{\rm OSL}$\tablefootmark{$\dag\dag$} \\
     (1) & (2)  & (3) & (4)  & (5)  & (6)  & (7) & (8) & (9)  & (10)\\
    \hline
    1 &  G121.31+0.64    &  00:36:47.80  &   63:28:57.0  &  0.89\tablefootmark{\rm 1}    &   78.0  &  0.052    & 22.8$\pm$0.1 &  0.15  & 0.12 \\
    2 &  G193.01+0.14    &  06:14:22.90  &   17:43:26.0  &  1.91\tablefootmark{\rm 1}    &  195.3  &  0.034    & 19.9$\pm$0.3 &  0.08  & 0.06 \\
    3 &      BGPS3604    &  18:30:43.92  &  -09:34:42.2  &  11.01\tablefootmark{\rm 2,3} &  114.8  &  0.050    & 12.3$\pm$0.4 &  0.26  & 0.21 \\
    4 &    G023.21$-$0.3   &  18:34:54.91  &  -08:49:19.2  &  4.60\tablefootmark{\rm 4,5}  &  192.8  &  0.037    & 24.0$\pm$0.2 &  0.57  & 0.45 \\
    5 &      BGPS4029    &  18:35:54.40  &  -07:59:44.6  &  3.54\tablefootmark{\rm 2,3}  &   78.4  &  0.048    & 14.1$\pm$0.2 &  0.29  & 0.23 \\
    6 &IRAS\,18360$-$0537  &  18:38:40.74  &  -05:35:04.0  &  6.30\tablefootmark{\rm 6,7}  &   78.4  &  0.054    & 28.0$\pm$0.2 &  0.33  & 0.26 \\
    7 &      BGPS5021    &  18:44:37.07  &  -02:55:04.4  &  5.18\tablefootmark{\rm 2,3}  &   78.1  &  0.048    & 14.7$\pm$0.2 &  0.41  & 0.32 \\
    8 &  G029.60$-$0.63    &  18:47:32.10  &  -03:14:10.0  &  4.31\tablefootmark{\rm 1}    &   78.4  &  0.056    & 16.7$\pm$0.2 &  0.23  & 0.18 \\
    9 &   G031.41+0.31   &  18:47:34.30  &  -01:12:46.0  &  3.75\tablefootmark{\rm 8}    &   77.8  &  0.065    & 40.2$\pm$0.8 &  0.20  & 0.15 \\
   10 &    G034.26+0.2   &  18:53:18.49  &   01:14:58.7  &  1.60\tablefootmark{\rm 4,5}  &   76.3  &  0.058    & 46.7$\pm$0.6 &  0.19  & 0.15 \\
   11 &    G035.20$-$0.7   &  18:58:12.93  &   01:40:40.6  &  2.20\tablefootmark{\rm 4,5}  &   96.4  &  0.078    & 27.3$\pm$0.2 &  0.60  & 0.47 \\
   12 &  G053.13+0.09    &  19:29:17.20  &   17:56:18.0  &  1.67\tablefootmark{\rm 1}    &  125.7  &  0.042    & 21.3$\pm$0.2 &  0.08  & 0.06 \\
   13 &          B335    &  19:37:00.43  &   07:34:06.8  &  0.16\tablefootmark{\rm 9}    &   95.7  &  0.060    & 12.8$\pm$0.2 &  0.41  & 0.32 \\
   14 &  G081.72+0.57    &  20:39:01.00  &   42:22:56.0  &  1.50\tablefootmark{\rm 1}    &   77.8  &  0.053    & 33.2$\pm$0.2 &  0.13  & 0.10 \\
   15 &  G082.21$-$1.53    &  20:49:33.00  &   41:27:34.0  &  1.13\tablefootmark{\rm 1}    &   97.0  &  0.047    & 11.2$\pm$0.3 &  0.42  & 0.33 \\
    \hline
    \raggedright
    \end{tabular}
    %\tablefoot{}
    \tablefoot{\tablefoottext{$\dag$} {The 1-$\sigma$ baseline noise level on a $T_{\rm mb}$ scale for a channel width of about 0.06 $\kms$.} \tablefoottext{$\dag\dag$} {The peak opacities of the inner ($\tau_{\rm ISL}$) and outer ($\tau_{\rm OSL}$) satellites.}
    \tablefoottext{\rm1} {\citet{2021ApJ...922..144Y}} \tablefoottext{\rm2}{\citet{2016ApJ...822...59S}} \tablefoottext{\rm 3}{\citet{2018ApJ...862...63C}} \tablefoottext{\rm 4}{\citet{2016A&A...585A.149W}} \tablefoottext{\rm 5}{\citet{2017A&A...599A.139K}} \tablefoottext{\rm 6}{\citet{2021ApJS..253....1X}} \tablefoottext{\rm 7}{\citet{2023A&A...677A..80W}} \tablefoottext{\rm 8}{\citet{2019A&A...632A.123I}} \tablefoottext{\rm 9}{\citet{2020RNAAS...4...88W}}.
    }
    \end{table*}

\section{Results}
\label{results}

    \begin{figure*}[t]
    \centering
    \includegraphics[width=0.33\hsize]{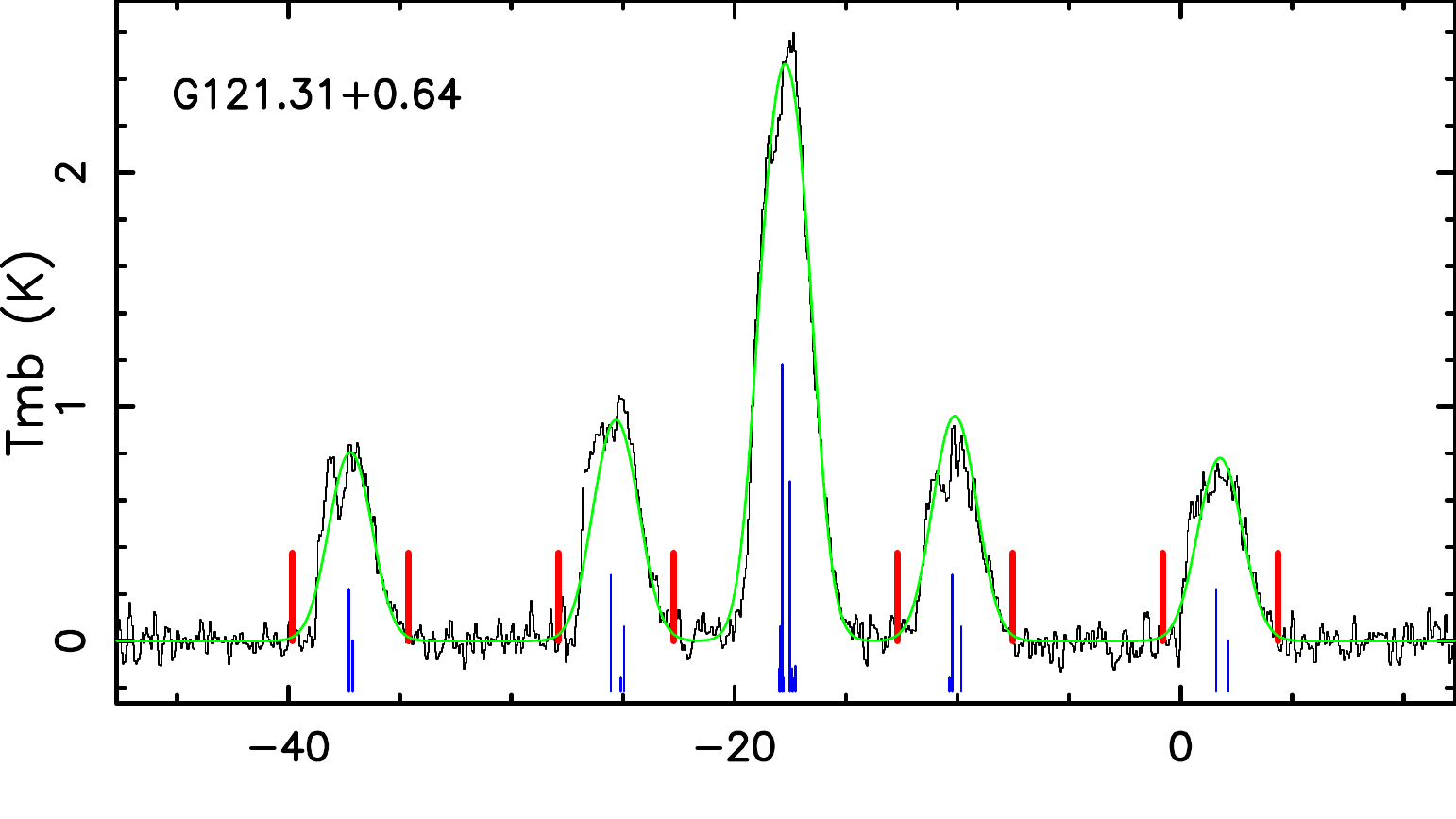}
    \includegraphics[width=0.33\hsize]{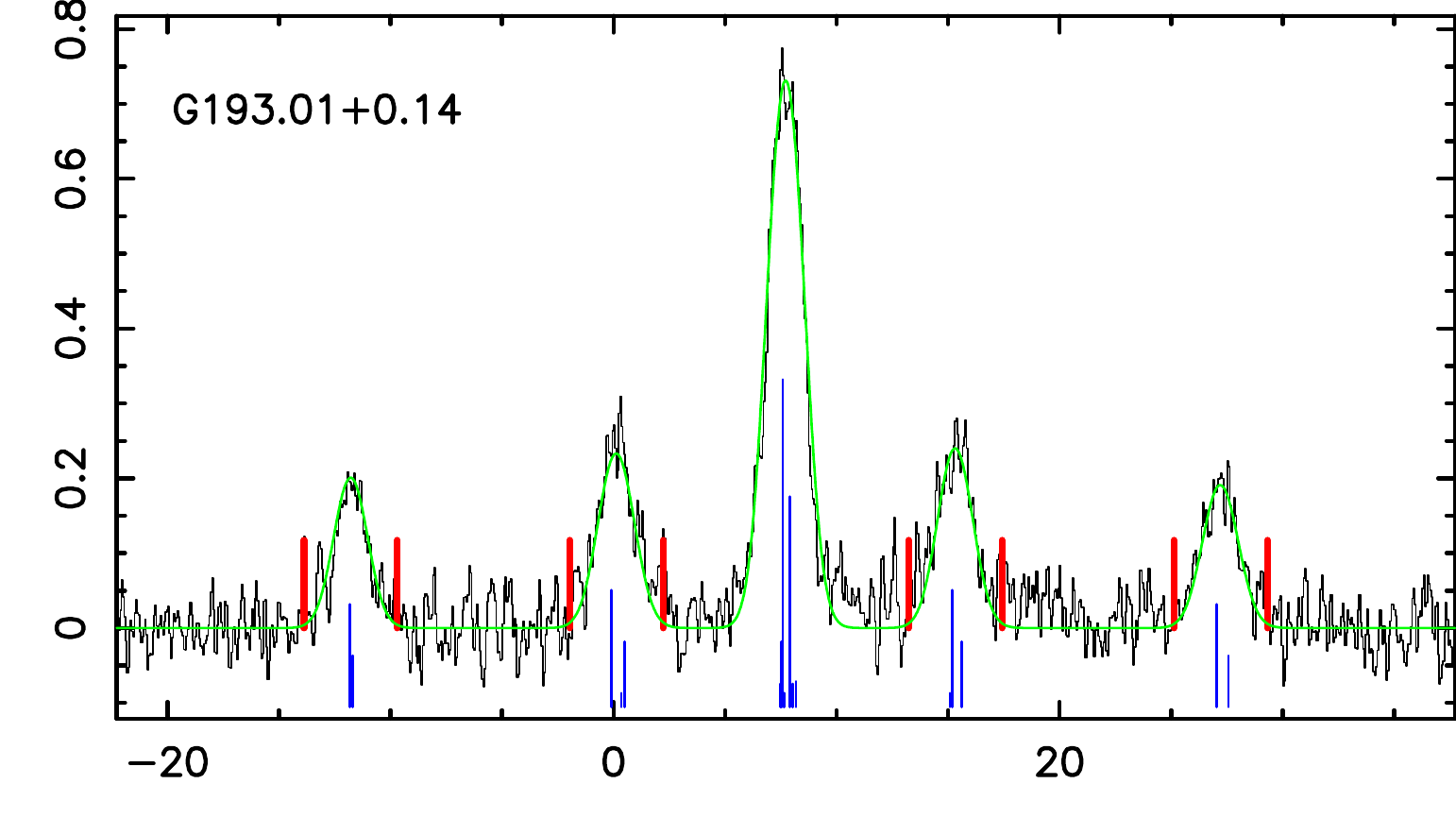}
    \includegraphics[width=0.33\hsize]{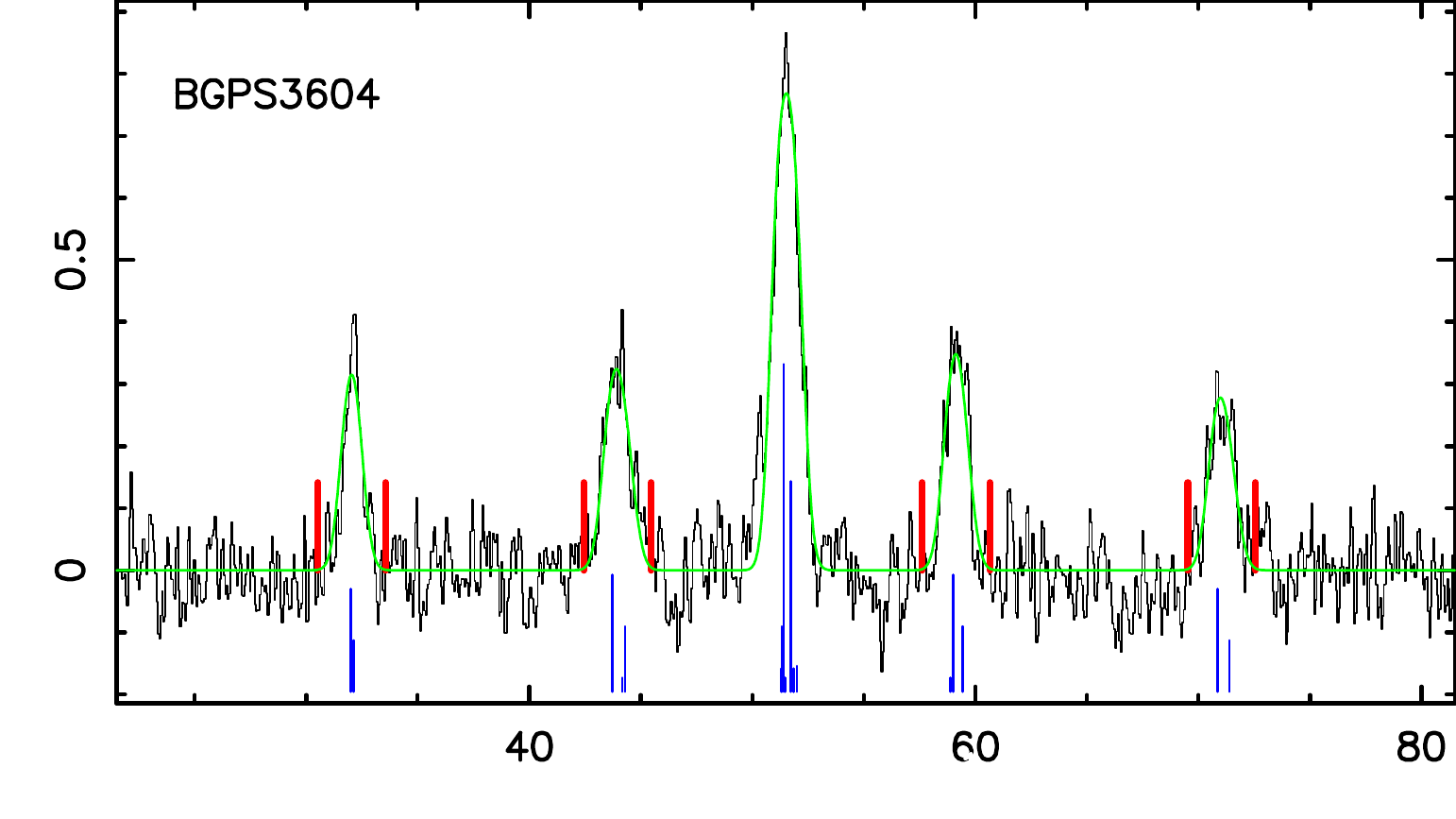}
    \includegraphics[width=0.33\hsize]{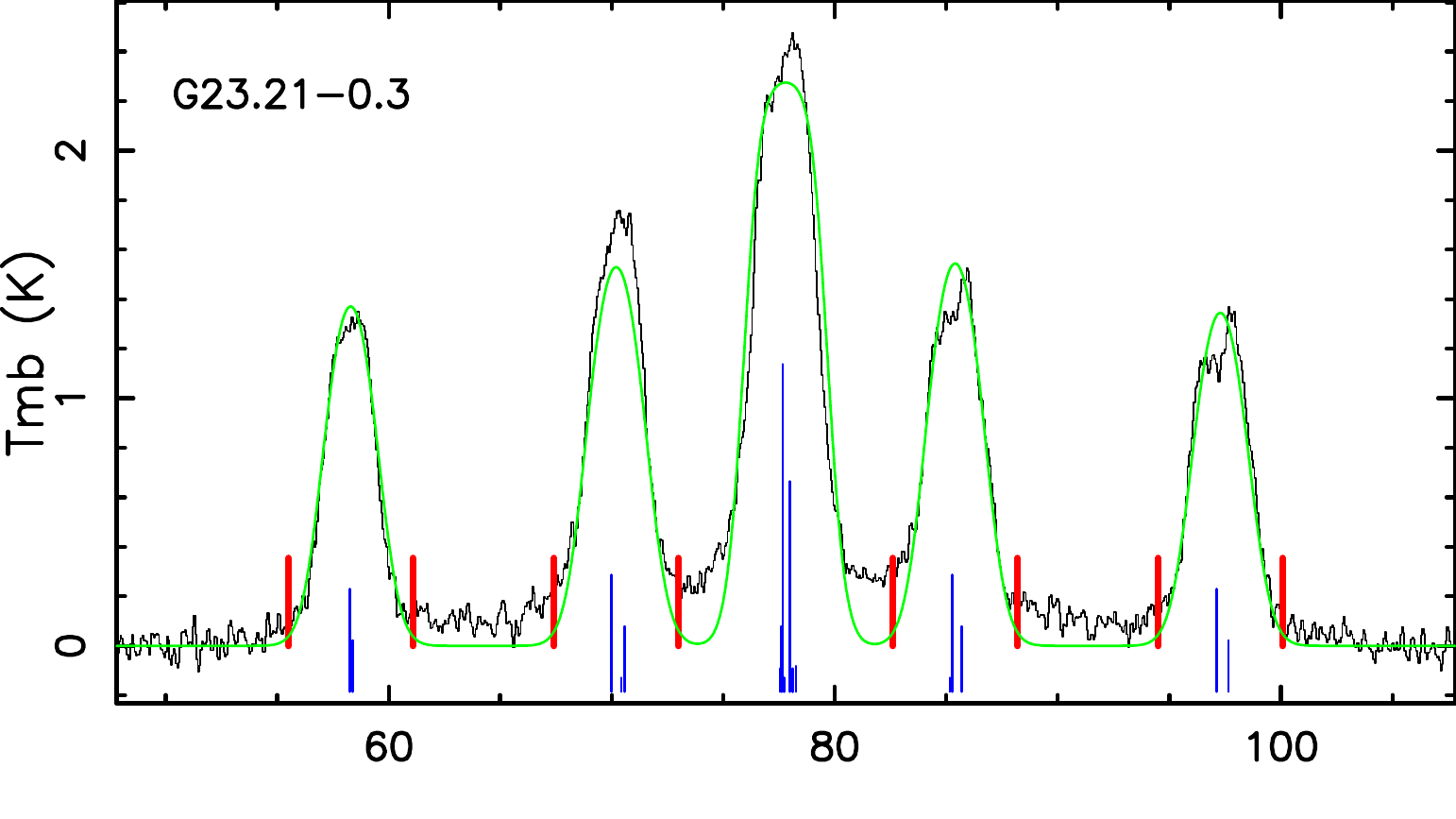}
    \includegraphics[width=0.33\hsize]{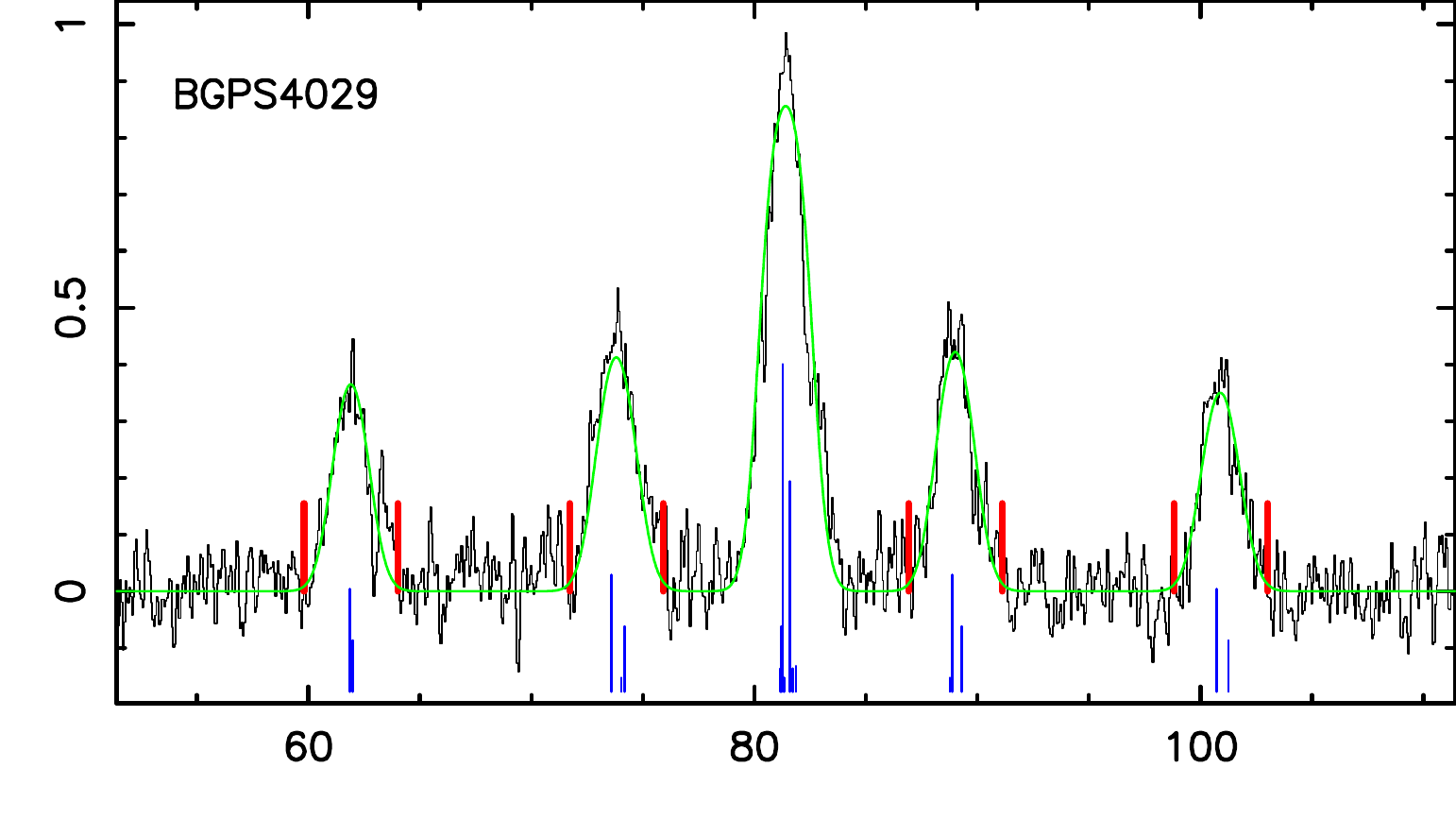}
    \includegraphics[width=0.33\hsize]{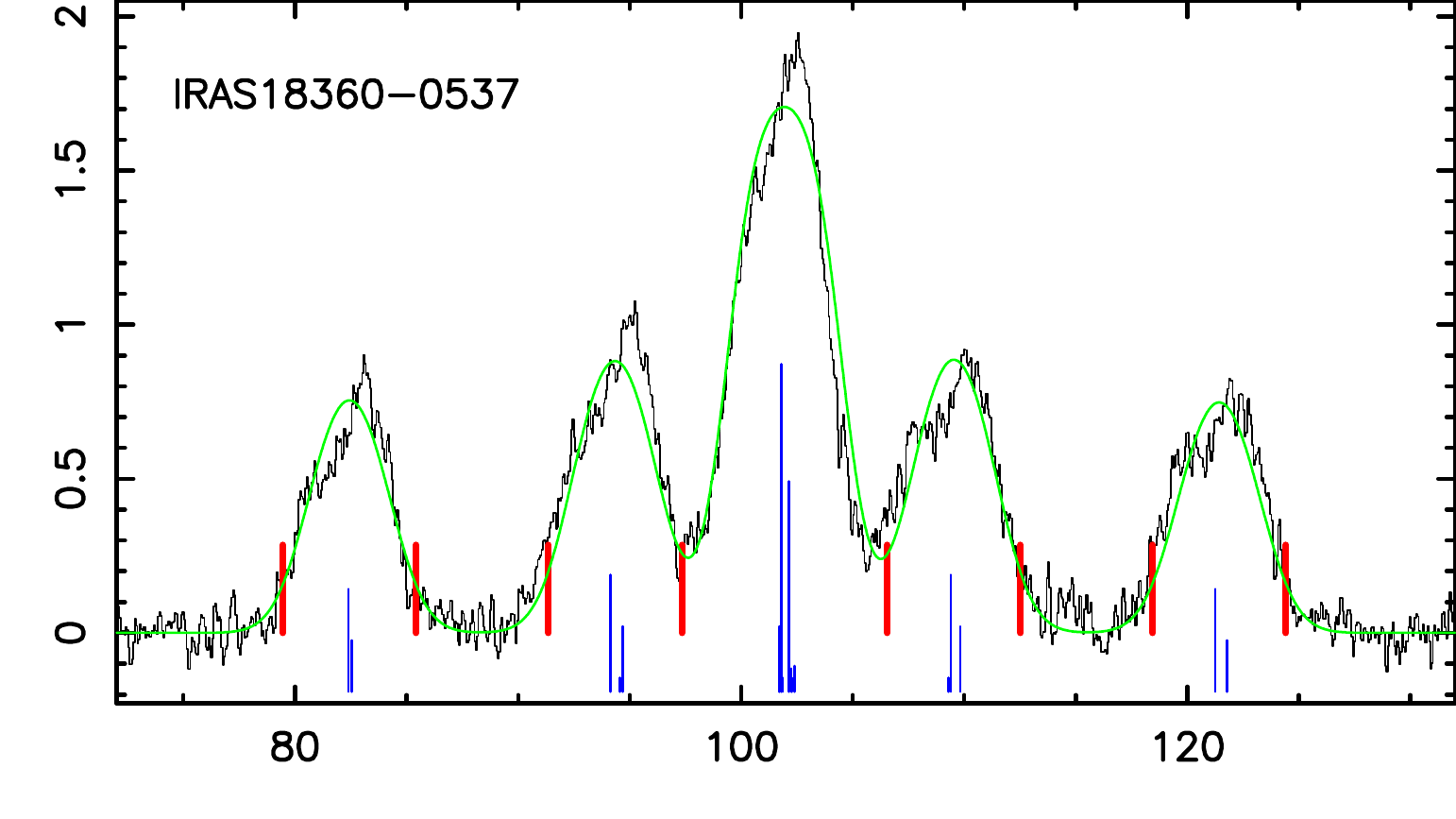}
    \includegraphics[width=0.33\hsize]{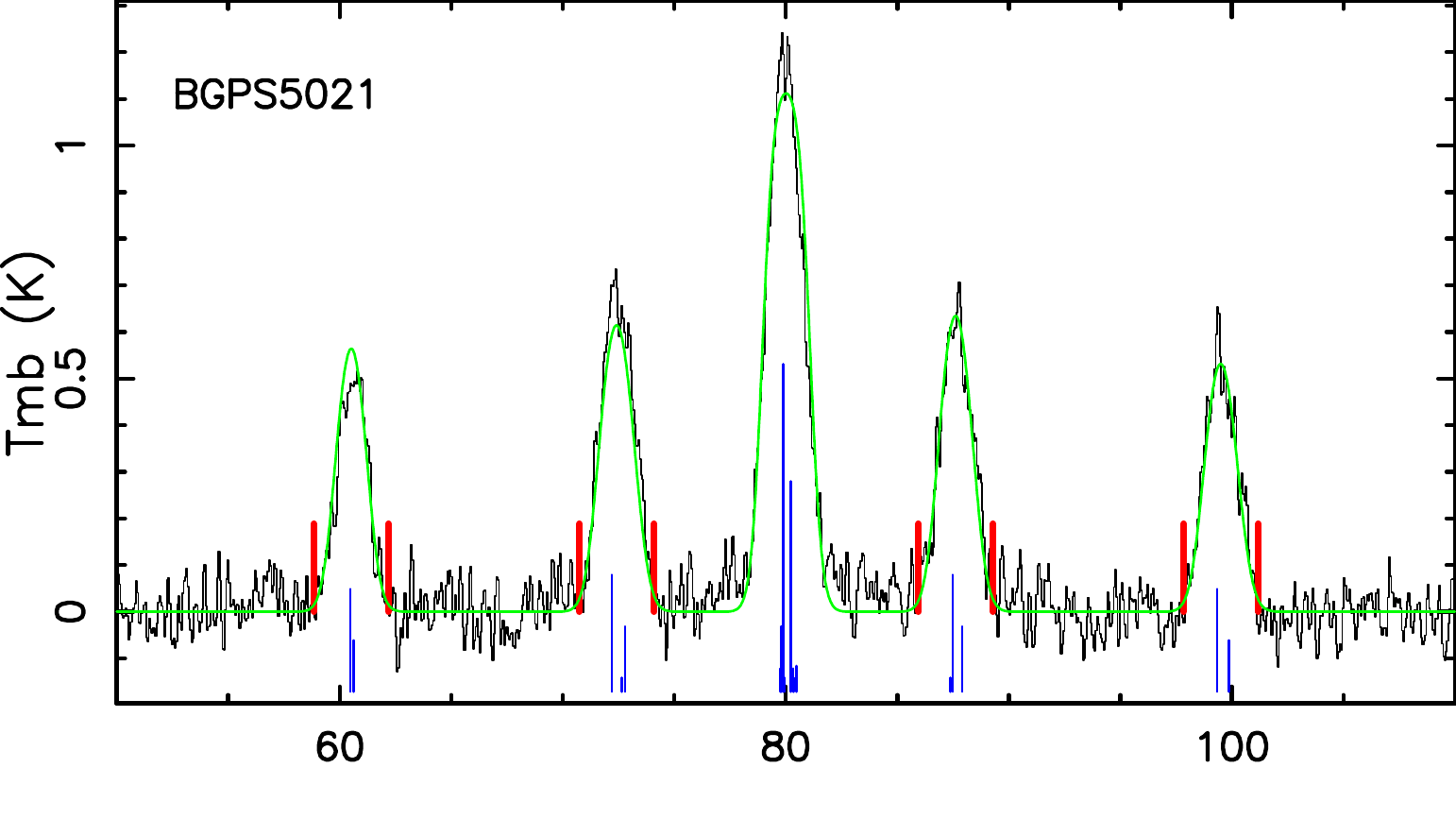}
    \includegraphics[width=0.33\hsize]{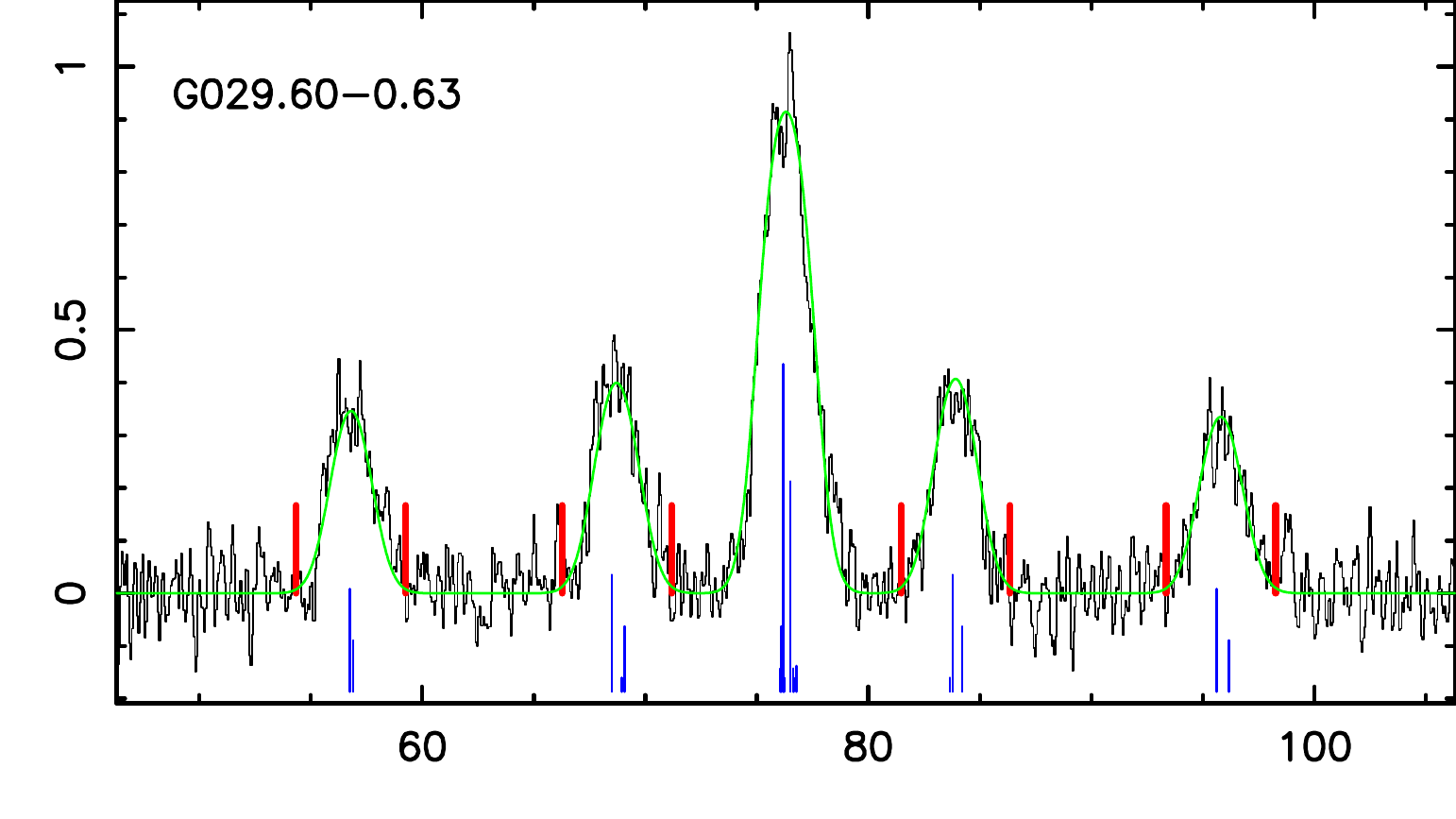}
    \includegraphics[width=0.33\hsize]{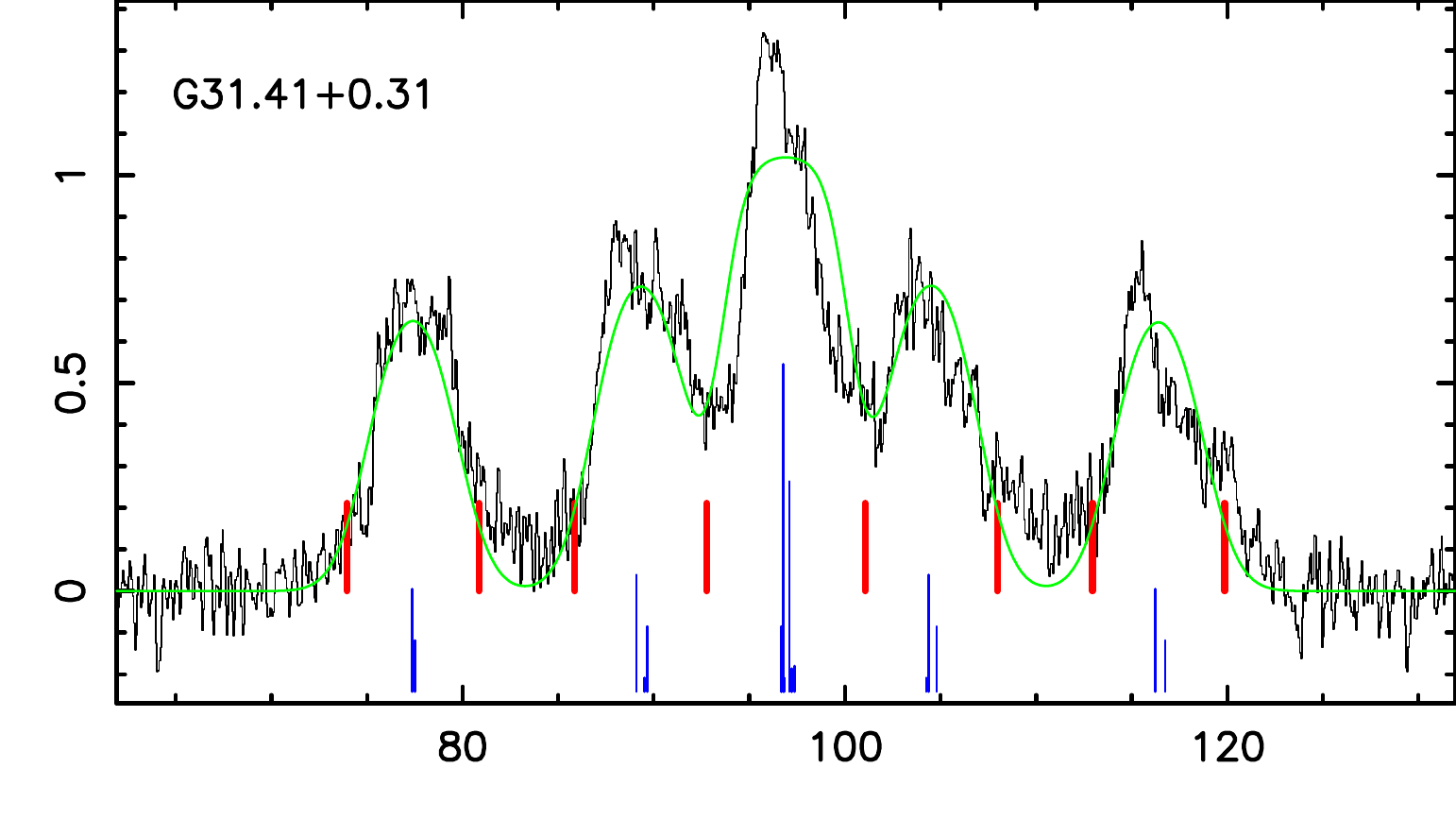}
    \includegraphics[width=0.33\hsize]{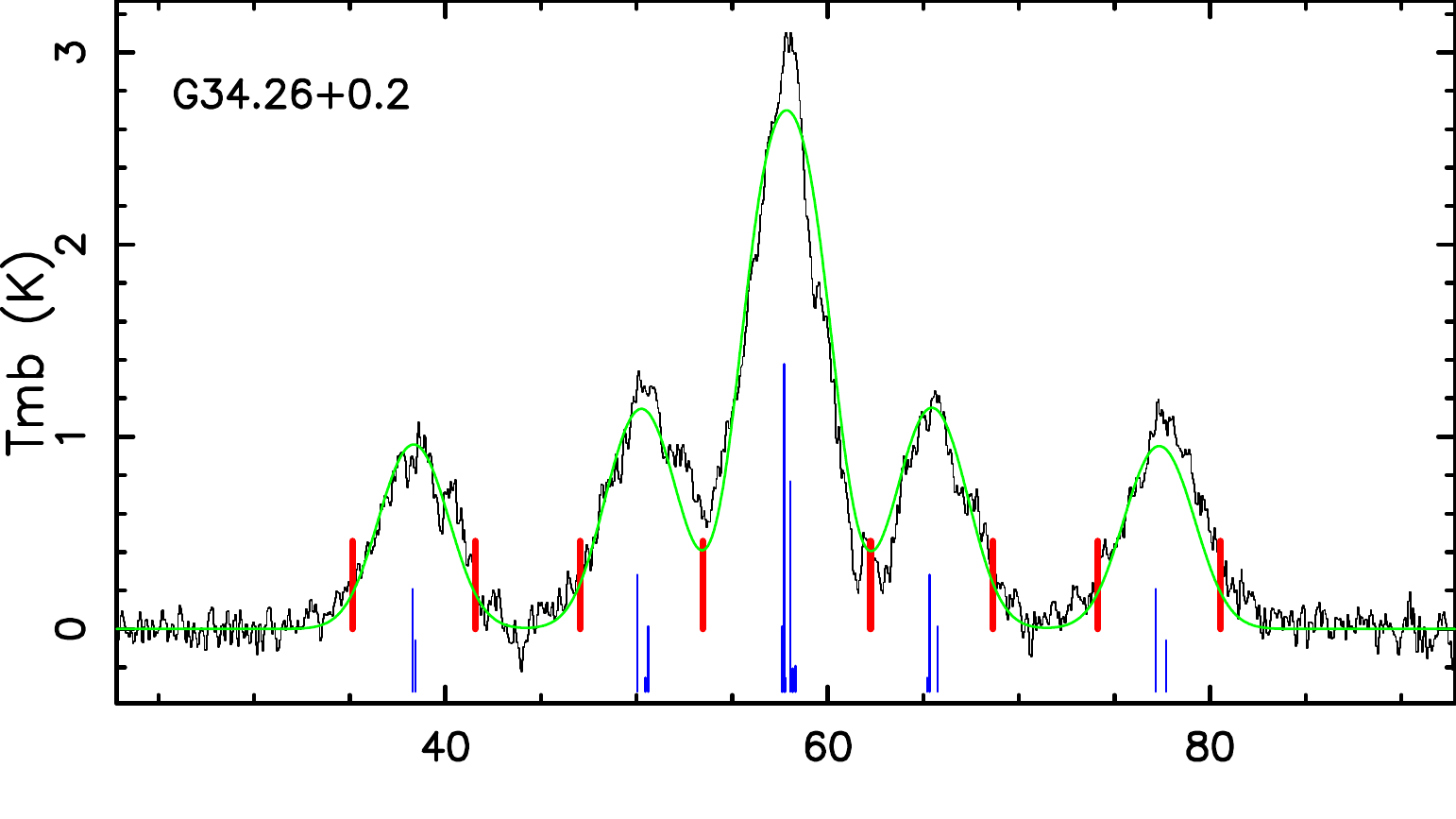}
    \includegraphics[width=0.33\hsize]{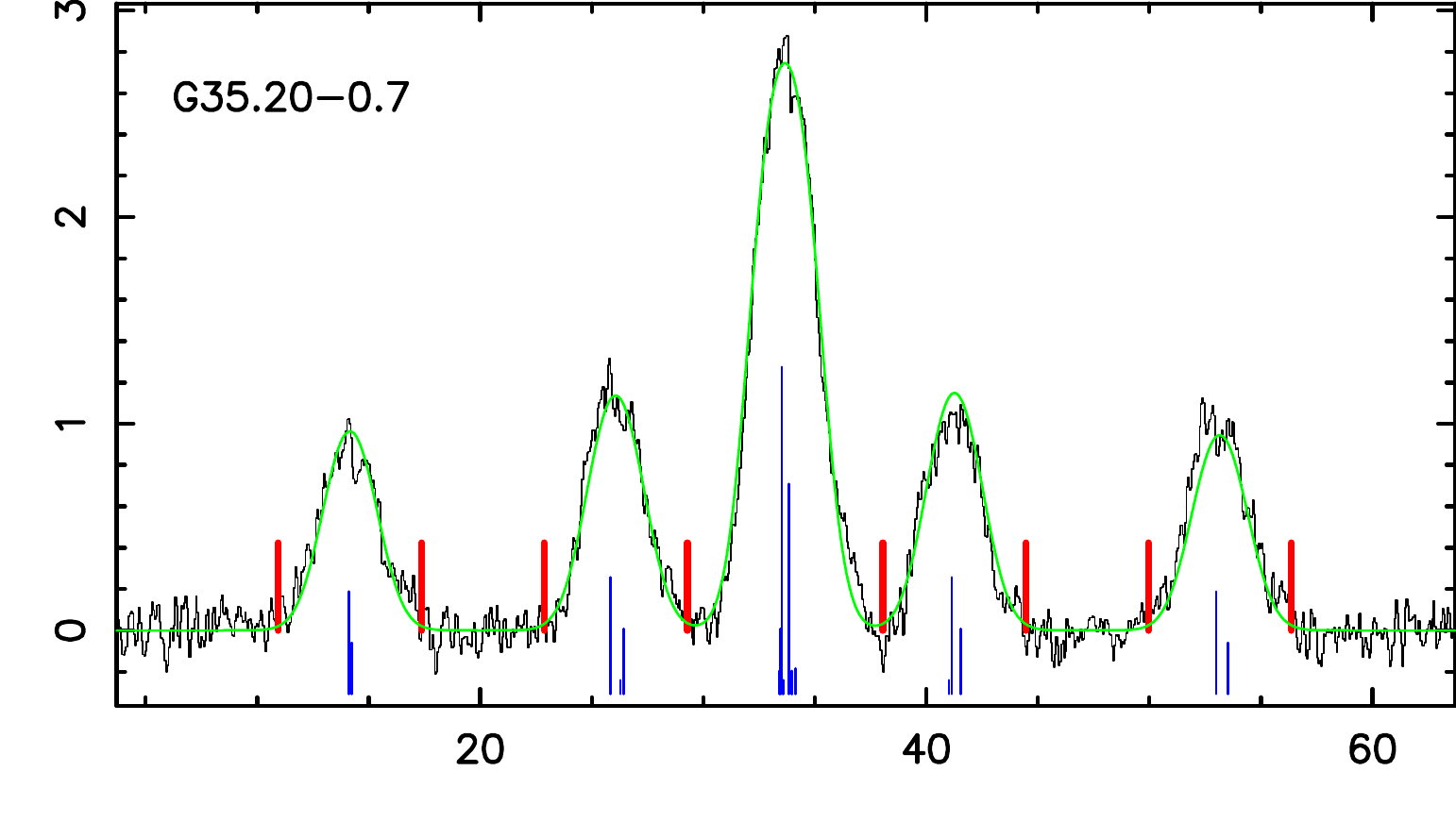}
    \includegraphics[width=0.33\hsize]{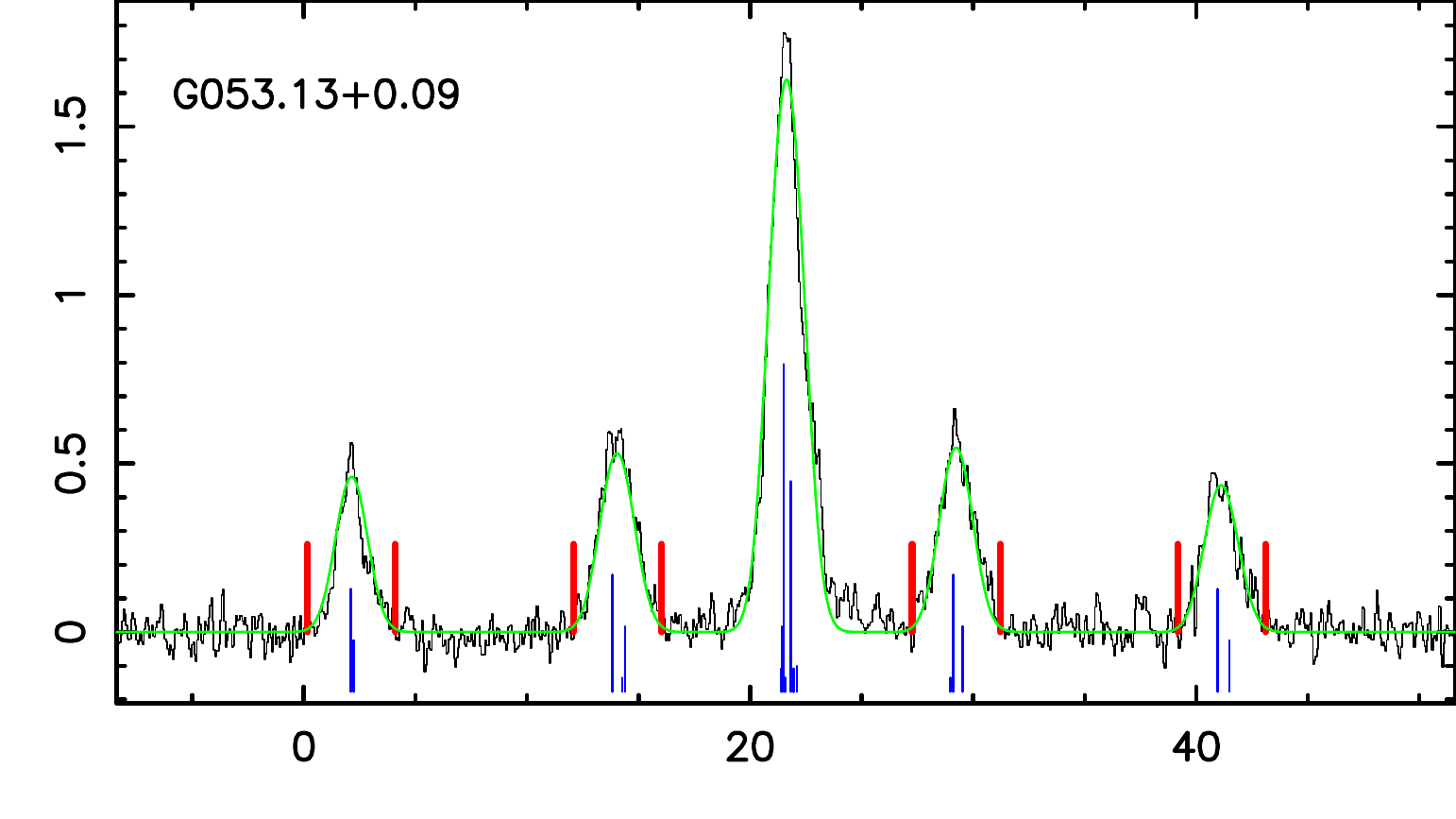}
    \includegraphics[width=0.33\hsize]{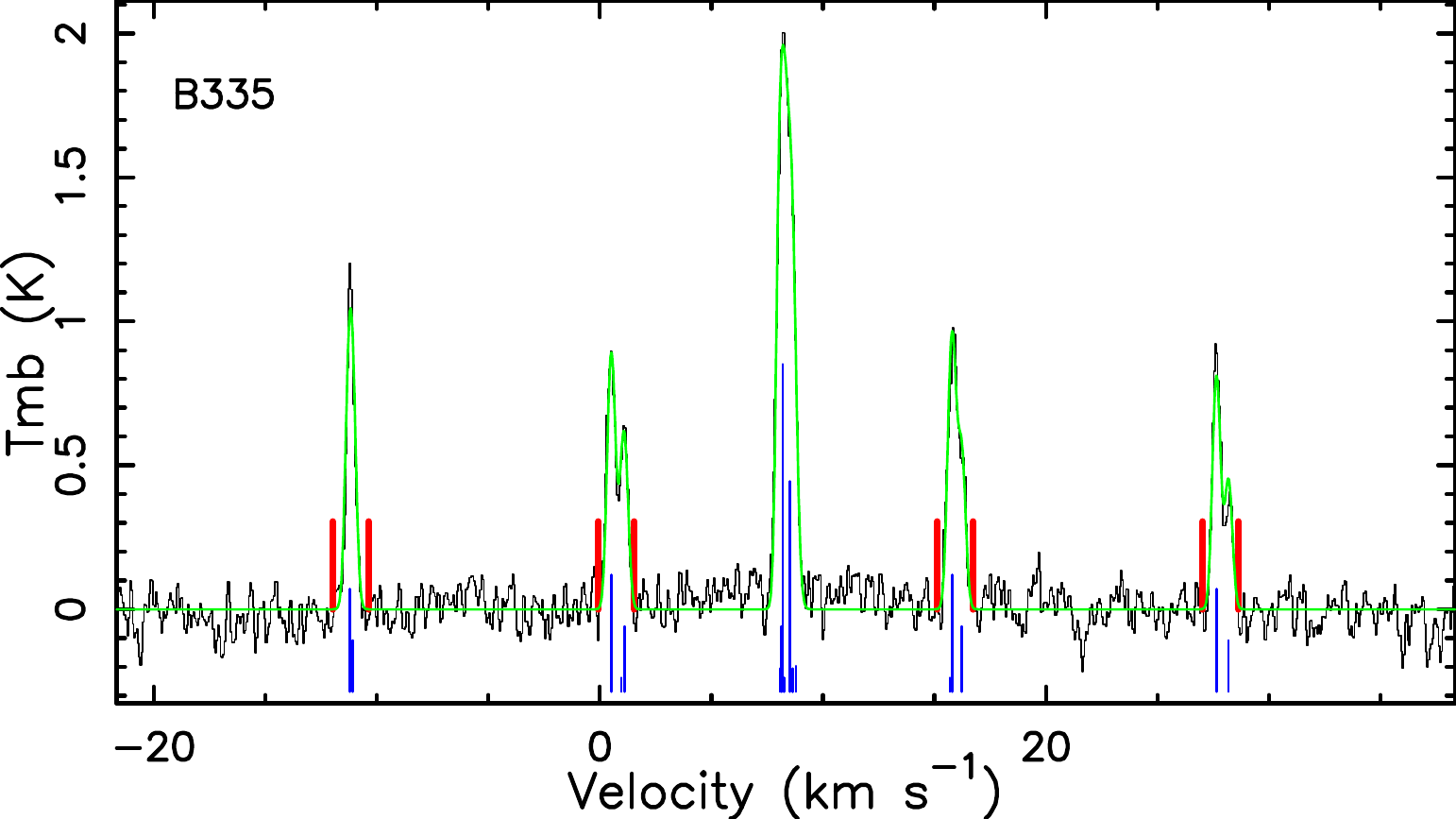}
    \includegraphics[width=0.33\hsize]{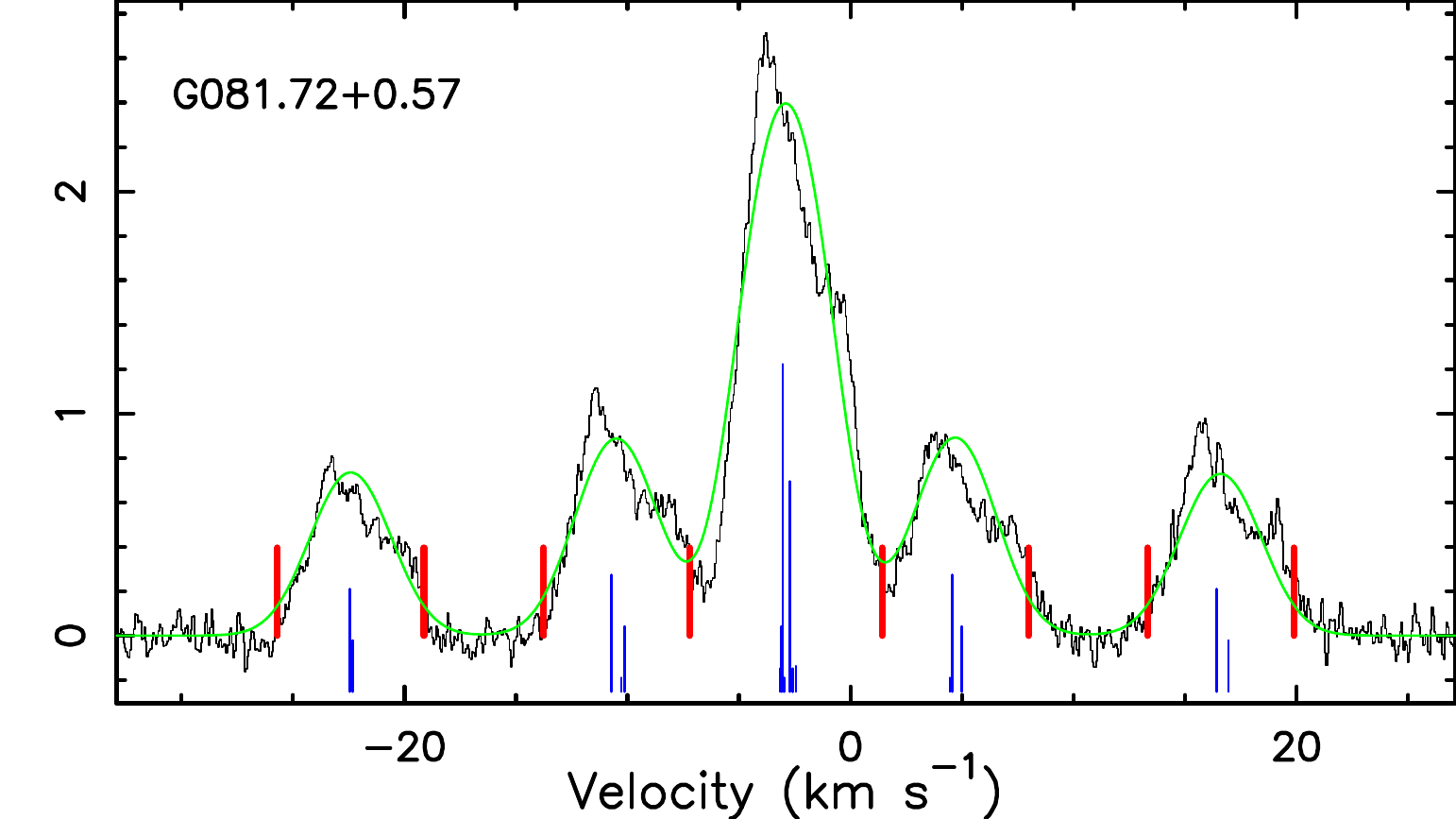}
    \includegraphics[width=0.33\hsize]{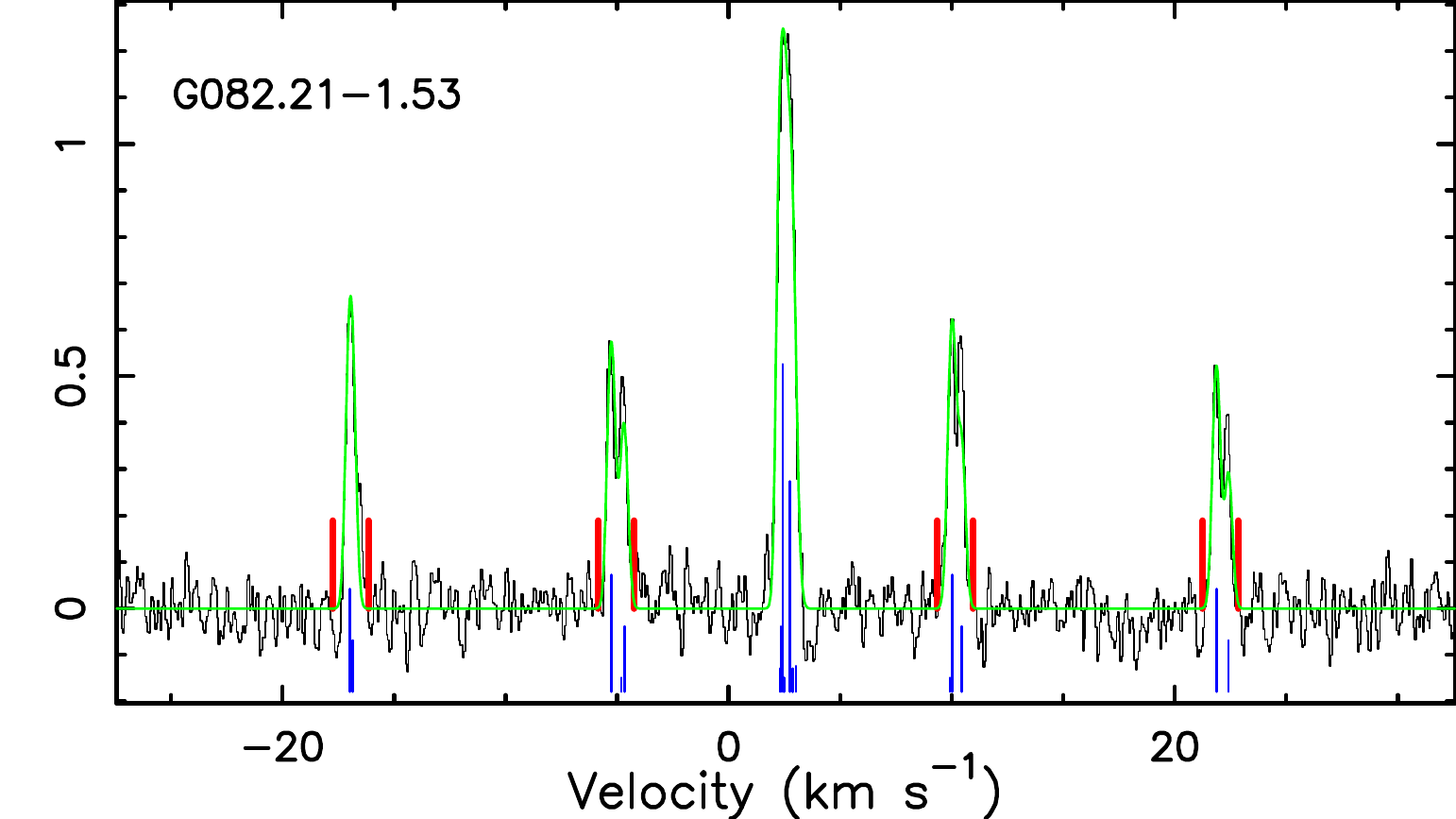}
    \caption{Observed NH$_{3}$ $(J,\,K) = (1,1)$ spectra. In each panel, the green curve represents the 18-hyperfines fitting result and the blue vertical lines under the spectrum indicate the position of the 18 hyperfine lines and their relative strengths in the optically thin case under conditions of LTE. The red lines denote the integrated ranges used to calculate the HIA. The source name is labeled in the top-left corner of each panel.}
    \label{fig:spec}
    \end{figure*}

Thanks to the long integration times on the targets (see Table \ref{tab:obs}), all of the NH$_{3}$ satellite lines are clearly detected (see Fig. \ref{fig:spec}).
This is important for unambiguously determining the HIAs and distinguishing between the HIA models (see Section \ref{intro} and below).
As discussed in \citet[][]{2020A&A...640A.114Z}, the HIA calculated with  peak intensities from Gaussian fittings does not accurately reflect the true anomaly.
That is because the red- and blueshifted ISLs (OSLs), which are the combination of Gaussian spectra of three (or two) hyperfine
components with different offsets (see the blue vertical lines in Fig. \ref{fig:spec}), should exhibit different line widths and peak intensities even under LTE and optically thin conditions \citep[see Appendix A in ][]{2020A&A...640A.114Z}.
We largely follow the recipe described in \citet[][]{2020A&A...640A.114Z} to determine the HIA by the ratio of their red- to blueshifted integrated intensities.
Specifically, we first used the combined 18 Gaussian hyperfine components to fit the observed NH$_{3}$ $(1,1)$ spectra.
Then, based on the fitted central velocity and velocity dispersion, we defined the integrated velocity ranges (see the red vertical lines in Fig. \ref{fig:spec}).
Finally, we calculated the HIAs of the ISL ($HIA_{\rm IS}$) and  OSL ($HIA_{\rm OS}$)  by the ratio of their redshifted  to blueshifted integrated intensities from the observed spectra,
     \begin{equation}
             HIA_{\rm IS} =  \frac{F_{\rm RISL}}{F_{\rm BISL}},\\
      \end{equation}
     \begin{equation}
        HIA_{\rm OS} = \frac{F_{\rm ROSL}}{F_{\rm BOSL}},\\
      \end{equation}
where $F_{\rm RISL}$/$F_{\rm ROSL}$ and $F_{\rm BISL}$/$F_{\rm BOSL}$ are the integrated intensities of the red-  and blueshifted sides of the ISLs/OSLs, respectively.
The standard deviation $\sigma_{\rm HIA}$ of HIAs is assigned by
     \begin{equation}
        \sigma_{\rm HIA} = HIA \times \sigma_{\rm BL} \times \sqrt{ N_{\rm C}/(F_{\rm R})^{2} + N_{\rm C}/(F_{\rm B})^{2} },\\
      \end{equation}
where HIA is either the value of $HIA_{\rm IS}$ or $HIA_{\rm OS}$.  $N_{\rm C}$ is the channel number within the integrated range and $\sigma_{\rm BL}$ is the noise level of the baseline. $F_{\rm R}$ and $F_{\rm B}$ are the integrated intensities of the redshifted and blueshifted sides of the ISLs or OSLs.

\begin{figure}[h]
    \includegraphics[width=\hsize]{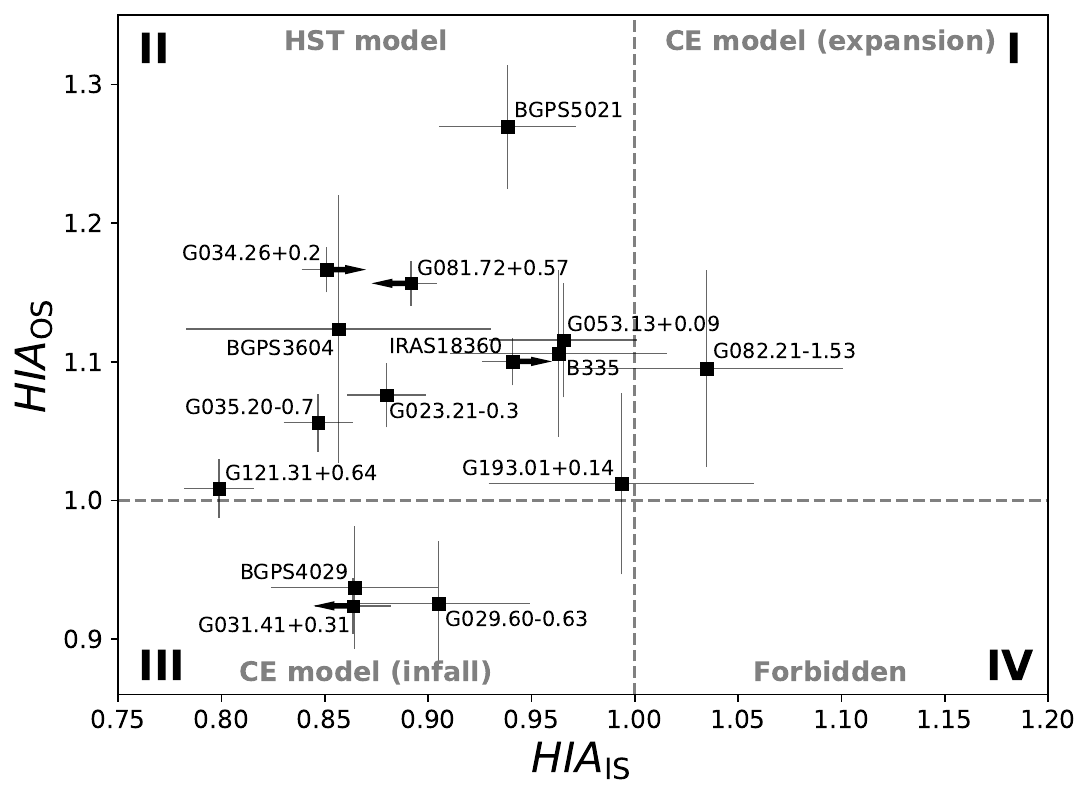}
    \centering
    \caption{Distribution of the hyperfine intensity anomalies of the inner ($HIA_{\rm IS}$) and outer ($HIA_{\rm OS}$ ) satellite lines. Arrows indicate potential corrections for the blending of the primary and inner satellite lines. The source name is marked in close proximity to each data point. Gray dashed vertical and horizontal lines divide the panel into four subregions: I for $HIA_{\rm IS}$>1 and $HIA_{\rm OS}$>1, II for $HIA_{\rm IS}$<1 and $HIA_{\rm OS}$>1, III for $HIA_{\rm IS}$<1 and $HIA_{\rm OS}$<1, and IV for $HIA_{\rm IS}$>1 and $HIA_{\rm OS}$<1. The models that cause HIAs to be located in these subregions are also labeled.}
    \label{fig:hia}
    \end{figure}

The distribution of observed $HIA_{\rm IS}$ and $HIA_{\rm OS}$ values of the 15 targets is also shown in Fig. \ref{fig:hia}.
Unity indicates no anomaly.
In 14 out of 15 targets, either $HIA_{\rm IS}$ or $HIA_{\rm OS}$ deviate from unity by more than $\sigma_{\rm HIA}$, and in 10 of these targets, both $HIA_{\rm IS}$ and $HIA_{\rm OS}$ values exceed $\sigma_{\rm HIA}$. Thus the presence of HIAs is prevalent in our sample.
The two dashed lines in Fig. \ref{fig:hia} divide the the $HIA_{\rm IS}$ and $HIA_{\rm OS}$ data into four quadrants: quadrants I ($HIA_{\rm IS}>1$ and $HIA_{\rm OS}>1$), II ($HIA_{\rm IS}<1$ and $HIA_{\rm OS}>1$), III ($HIA_{\rm IS}<1$ and $HIA_{\rm OS}<1$), and IV ($HIA_{\rm IS}>1$ and $HIA_{\rm OS}<1$).
These quadrants can be utilized to distinguish different HIA models (see Section \ref{hia-infall}).
From Fig. \ref{fig:hia} we can see that 1 (6.7\%), 11 (73.3\%), 3 (20\%), and 0 data points are located in quadrants I, II, III, and IV (0, 7 (46.7\%), 3 (20\%), and 0, respectively, considering 1-$\sigma$ uncertainties).
These fractions in the four quadrants, based on our deep observations, are generally consistent with previous statistical results \citep[e.g.][]{2018A&A...609A.125W,2020A&A...640A.114Z}.

In the cases of IRAS\,18360-0537, G031.41+0.31, G034.26+0.2, and G081.72+0.57, there is significant blending between their main lines and ISLs.
This may impact the determination of the $HIA_{\rm IS}$ if the main line has asymmetric profiles, which contributes differently to the two ISLs.
Hereby, we qualitatively discuss the potential corrections for the line blending.
Taking IRAS\,18360-0537 as an example, as seen in the first panel of Fig. \ref{fig:spec} that the asymmetric main line contributes more flux to the blueshifted ISL than to the redshifted ISL.
Thus the real $HIA_{\rm IS}$ might be larger than the calculated one, as indicated by the arrow in Fig. \ref{fig:hia}.
However, we should note that the length of the arrow is not quantitatively proportional to the corrections, as it is difficult to precisely determine the contribution of the blending main line.
For all of these four targets, we labeled the potential corrections with arrows in Fig. \ref{fig:hia}.
We can see that, for G031.41+0.31 and G081.72+0.57, their actual $HIA_{\rm IS}$ should be smaller than the calculated one.
Thus these corrections do not alter their quadrants.
For the other two targets, especially IRAS\,18360-0537 whose $HIA_{\rm IS}$ is closer to unity, the line blending effect might change their quadrant from II to I.

We should note that changes in the relative intensities of the ISL or OSL may occur due to large opacities.
%
%Within each  ISL (OSL), there are three (two) hyperfine components exhibiting identical intensities \textbf{denoted by the heights of blue vertical lines in Fig. \ref{fig:spec}}. Nonetheless, as previously noted, the frequency separations among hyperfine components are different \textbf{denoted by the separations of blue vertical lines in Fig. \ref{fig:spec}} \citep[see also e.g.][]{1977ApJ...215L..35R}.
%
Theoretically, each ISL (OSL) contains three (two) identical hyperfine components, indicated by the blue vertical lines in Fig. \ref{fig:spec}, However, as previously mentioned, while the intensities of the hyperfine components are the same, the frequency separations between them vary, as represented by the spacing of the blue vertical lines in Fig. \ref{fig:spec}.
In high-opacity scenarios, satellites with closer separations between hyperfine components may achieve saturation more rapidly compared to those with larger separations.
Therefore, we also derived the opacities of the ISL and OSL for all the spectra by summing up the opacities of their respective hyperfine components (see Table \ref{tab:obs}). The ISLs and OSLs in our spectra tend to be optically thin, with the exceptions of those of G035.2$-$0.7 and G023.21$-$0.3, which show moderate opacities of about 0.6.
Therefore, the opacities should not result in a significant intensity difference between the two ISLs/OSLs.

As discussed in \citet{2020A&A...640A.114Z}, the deviations of the HIA defined by peak intensities from the true HIA is getting more pronounced for spectra with narrow linewithds.
We can see from Fig. \ref{fig:hia-isos}, take B335 as an example, that the peak intensity of the blueshifted OSL is larger than that of the redshifted one,
just because the separation of the two hyperfine components within the blueshifted OSL is smaller than that of the redshifted OSL (see the blue vertical lines in Fig. \ref{fig:hia}).
From Fig. \ref{fig:hia} we see that $HIA_{\rm OS}$ of B335 is actually larger than unity (the integrated intensity of the blueshifted OSL is smaller than that of the redshifted one).
We refer to the Appendix A in \citet{2020A&A...640A.114Z} for the detailed comparisons between the HIAs defined by peak and integrated intensity ratios.
In addition, at the resolution of 37$\as$, our observations may encompass a number of cloud or velocity components, which are more evident in the case of G31.41+0.31 and IRAS\,18360$-$0537, which show slight deviations from Gaussian profiles (see Fig. \ref{fig:spec}). While the limited angular resolution of our data is ideal for the determination of the gross HIA, it does not permit an evaluation of the spatial fine structure, eventually revealing more than one velocity component within the targeted area.

\section{Discussion}

\subsection{Are HIAs sensitive to infall motions?}
\label{hia-infall}

The HST and CE models predict distinct HIAs \citep[e.g.][]{2020A&A...640A.114Z}.
According to the HST model, the intensity of the blueshifted ISL is expected to be stronger than that of the redshifted one, while the redshifted OSL should exhibit stronger intensity than the blueshifted one.
Thus $HIA_{\rm IS}<1$ and $HIA_{\rm OS}>1$ \citep[quadrant II; e.g.][]{1985A&A...144...13S, 2015ApJ...806...74C}.
In the context of the CE model \citep[e.g.][]{2001A&A...376..348P}, for infall motions, both the blueshifted ISL and OSL are anticipated to be stronger than the redshifted two lines, that is $HIA_{\rm IS}<1$ and $HIA_{\rm OS}<1$ (quadrant III). Conversely, for expansion motions, the redshifted ISL and OSL are expected to exhibit stronger intensity simultaneously, that is $HIA_{\rm IS}>1$ and $HIA_{\rm OS}>1$ (quadrant I).
As discussed in \citet{2020A&A...640A.114Z}, we can employ the HIA quadrant plot of $HIA_{\rm IS}$ and $HIA_{\rm OS}$ to identify the HIA models (see the different models labeled in the four subregions of Fig. \ref{fig:hia}).

From Fig. \ref{fig:hia}, it is noteworthy that all the derived HIAs remain within the framework of the two models, with no data points in the forbidden quadrant IV.
However, it is somewhat unexpected that a majority of the HIAs are located in the second quadrant, aligning with the HST model.
This indicates that, in our observations, the HST model is the predominant model, even in sources likely harboring infall motions.
However, alternative explanations exist. Similar to the case of blue-skewed profiles, they might be blended with emission from outflows \citep[e.g.][]{2003cdsf.conf..157E, 2016A&A...585A.149W}.
This phenomenon is also applicable to HIAs, since ammonia emission is also commonly seen in outflows \citep[e.g.][]{2007A&A...470..269Z}.
This could only be further clarified through interferometer observations, which are capable of resolving distinct structures.

There are three data points located in quadrant III, indicating the potential existence of infall motions.
For these three sources, infall motions should be widespread within the beam size of about 37$\as$.
Increased kinetic temperatures would enhance the HST HIAs \citep[e.g.][]{1985A&A...144...13S} and could also lead to more ammonia molecules being excited to the NH$_{3}$ $(2,1)$ level.
Consequently, the likelihood of sources to be fit the HST model dominating is heightened.
So, HIAs may preferably serve as infall tracers in star-forming regions at early evolutionary stages.
However, we do not find a preferred evolutionary stage among these three targets in quadrant III.
BGPS\,4029, G029.60$-$0.63, and G031.41+0.31 were reported to be associated with an infrared dark cloud (IRDC) \citep[][]{2009A&A...505..405P}, Class 0/I YSOs \citep[][]{2020RAA....20..115Y}, and a massive protocluster, respectively \citep[][]{2022A&A...659A..81B}.
Nevertheless, HIAs may indeed be used as an infall tracer for early evolutionary stages, such as IRDCs.

As mentioned earlier, in the HST model, the ISL and OSL exhibit reversed anomalous intensities ($HIA_{\rm IS}<1$ and $HIA_{\rm OS}>1$). In the CE model, both the blueshifted/redshifted ISL and OSL should be weaker/stronger simultaneously for infall/expansion motions ($HIA_{\rm IS}<1$ and $HIA_{\rm OS}<1$ for infall motions, $HIA_{\rm IS}>1$ and $HIA_{\rm OS}>1$ for expansion motions).
As a result, using the ratio between the sum of the intensities of the two redshifted satellite lines and the sum of the intensities of the two blueshifted satellite lines, the influence of the HST model (the ISL and OSL exhibit reversed anomalies) would be mitigated and that of the CE model would be enhanced.
Therefore we also study the ratio $HIA_{\rm ISOS}$,
     \begin{equation}
             HIA_{\rm ISOS} =  \frac{F_{\rm RISL} + F_{\rm ROSL}}{F_{\rm BISL}+F_{\rm BOSL}}.\\
      \end{equation}
We should note that it is not known whether the anomalous fluxes of the blueshifted ISL and the redshifted OSL are precisely equal in the HST model \citep[e.g.][]{1985A&A...144...13S}.

\begin{figure}[h]
    \includegraphics[width=\hsize]{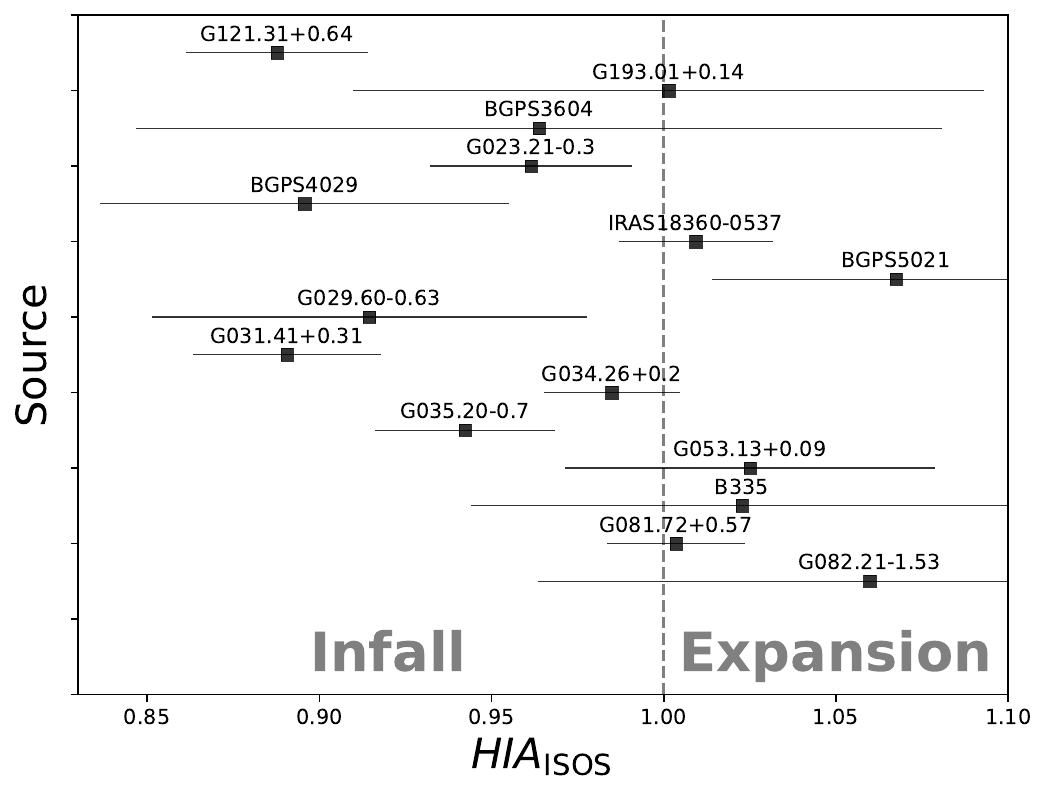}
    \centering
    \caption{Distribution of the combined hyperfine intensity anomalies. The source name is labeled above each data point.}
    \label{fig:hia-isos}
    \end{figure}

Fig. \ref{fig:hia-isos} shows the distribution of $HIA_{\rm ISOS}$.
Eight sources (six, considering uncertainties) out of 15 are consistent with infall motions.
Indeed, more sources exhibit consistency with infall motions than those in quadrant III of Fig. \ref{fig:hia}. %KMM Than WHAT????
This outcome demonstrates that  $HIA_{\rm ISOS}$ could be a better infall tracer than $HIA_{\rm IS}$ and $HIA_{\rm OS}$.
However, $HIA_{\rm ISOS}$ may also not be a very ideal indicator of infall for our data. Taking uncertainties into account, there are only six reliable data points of $HIA_{\rm ISOS}$  consistent with infall motions in this sample of infall candidates.
Naturally, the three sources already previously suggested to represent infall (quadrant III sources in Fig. \ref{fig:hia}) are part of this sub-sample.

To summarize, in our single-dish observations, most of the detected HIAs are consistent with the HST model.
HIAs could be used as an infall tracer but seem not highly sensitive to infall motions.
High-resolution observations would be essential for a more precise assessment of the contaminating impact by outflows.

\subsection{HIAs versus kinetic temperature}

\begin{figure*}[t]
    \includegraphics[width=\hsize]{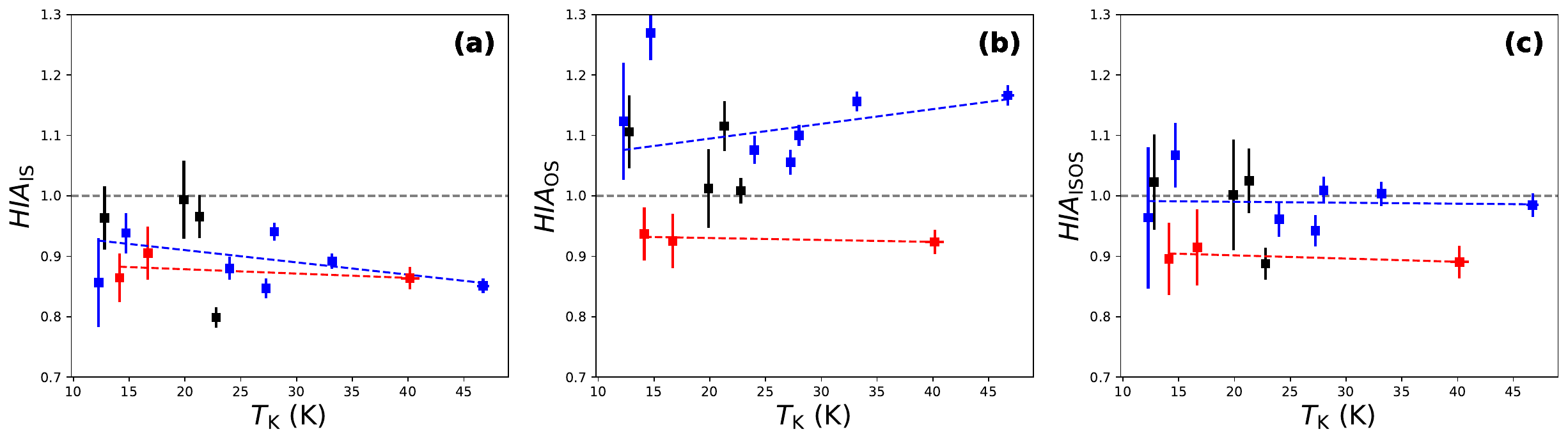}
    \centering
    \caption{Correlations of $HIA_{\rm IS}$ and $T_{\rm K}$ (panel a), $HIA_{\rm OS}$ and $T_{\rm K}$ (panel b), and $HIA_{\rm ISOS}$ and $T_{\rm K}$ (panel c). The HIAs consistent with the HST model and CE model (infall motions) are emphasized in blue and red colors, respectively. Blue and red lines indicate their linear regression results. Linear fit parameters for the blue and red lines are given in Table \ref{tab:hia_paras}.}
    \label{fig:hia-paras}
    \end{figure*}

As we mentioned before,  higher temperatures would potentially lead to the dominance of the HST model.
In this section, we explore the correlation between HIAs and the kinetic temperature ($T_{\rm K}$).
NH$_{3}$ inversion transitions are an invaluable spectroscopic probe of  $T_{\rm K}$ \citep{1983ARA&A..21..239H}.
We derive the rotational temperature $T_{\rm R}$ from the observed para-NH$_{3}$ $(1,1)$ and $(2,2)$ spectral lines (see Fig. \ref{fig:spec112233}) using the PySpecKit package \citep{2011ascl.soft09001G}, which is a forward-modeling tool for spectral lines.
Then, $T_{\rm K}$ is calculated following \citet{2004A&A...416..191T} as
     \begin{equation}
             T_{\rm K} =  \frac{T_{\rm R}}{1- \frac{T_{\rm R}}{42} \rm ln[1+1.1 exp(-16/\it T_{\rm R})]  }.\\
      \end{equation}
\citet{2004A&A...416..191T} conducted various Monte Carlo simulations involving data of the  NH$_{3}$ $(1,1), (2,1)$, and $(2,2)$ transitions to derive an analytical expression to estimate $T_{\rm K}$ from $T_{\rm R}$.
$T_{\rm K}$ can be very well approximated by this equation in the range $T_{\rm K}$ = 5--20\,K. It is important to note that the applicability of this approximation diminishes at higher temperatures.
Fig. \ref{fig:hia-paras} shows the correlations between HIAs ($HIA_{\rm IS}$,  $HIA_{\rm OS}$, and  $HIA_{\rm ISOS}$) and $T_{\rm K}$.
The HIAs consistent with the HST model and CE model (infall motions) are emphasized in blue and red colors, respectively.
Meanwhile, their linear regression results are also shown in each panel (blue and red lines) and in Table \ref{tab:hia_paras}.

We first see from the blue data points in panels (a) and (b) of Fig. \ref{fig:hia-paras} that $HIA_{\rm IS}$ and $HIA_{\rm OS}$ all tend to show rising deviations from unity (indicating higher anomalies) with increasing $T_{\rm K}$, which is expected by the HST model. We should note that the correlations are weak, with correlation coefficients of $-$0.29 and $-$0.12 for $HIA_{\rm IS}$ and $HIA_{\rm OS}$, respectively (see Table \ref{tab:hia_paras}).
This may be attributed to the considerable dispersion among these data points, especially $HIA_{\rm OS}$ determined by the relatively weak OSLs.
While the HST model predominantly influences the blue data points, the CE model may also play a minor role, introducing additional uncertainties. For example, the two HIA models produce contrasting predictions concerning the enhancement of the OSLs.

HIAs consistent with infall motions are not expected to exhibit a dependence on $T_{\rm K}$.
In our observations, both $HIA_{\rm IS}$ and $HIA_{\rm OS}$, indicated by red color, appear to show a constant value not depending on $T_{\rm K}$.
It is essential to note that the sample size is limited, consisting of only three data points.
Therefore, subsequent comparisons of potential trends in HIAs against $T_{\rm K}$ are tentative.
The slope of $HIA_{\rm IS}$ appears more negative compared to that of $HIA_{\rm OS}$.
In general, HIAs induced by infall motions (red data points) demonstrate relatively lower sensitivity to $T_{\rm K}$ compared to HIAs consistent with the HST model (blue data points in Fig. \ref{fig:hia-paras}).
This outcome implies that HIAs might serve as more effective infall tracers for relatively cold gas, which may help us to understand the large-scale accretion and infall motions in early evolutionary star-forming cores.

Finally, in panel (c) of Fig. \ref{fig:hia-paras}, we can see from the blue data points, in general, the $HIA_{\rm ISOS}$ of the targets associated with the HST model, are very close to unity.
So, $HIA_{\rm ISOS}$ could largely weaken the impact of the HST model as explained before (see Fig. \ref{fig:hia-isos}). This indicates that the anomalous flux to the blueshifted ISL is comparable to that of the redshifted OSL in the HST model.
%
%
%For the red data points \textbf{(i.e. those associated with infall motions)}, they are not influenced by the combination of the satellite lines, \textbf{since their blueshifted ISL and OSL are enhanced simultaneously.}
The red data points (i.e. those associated with infall motions) are not affected by such a consideration, since their blueshifted ISLs and OSLs are enhanced simultaneously.

\begin{table}
\caption{Parameters for the linear regressions.}
\label{tab:hia_paras}
\centering
%\raggedright
\begin{tabular}{cccc}
\hline\hline
 Models  & Slope ($\times$10$^{-3}$) & Intercept & r\tablefootmark{$\dag$} \\
 (1) & (2)  & (3)   \\
 \hline\hline
\multicolumn{4}{c}{$HIA_{\rm IS}$} \\
\hline
CE   &  -0.7$\pm$1.0    &  0.90$\pm$0.04  & -0.43\\
HST  &  -2.0$\pm$1.4    &   0.93$\pm$0.05 & -0.29\\
\hline
\multicolumn{4}{c}{$HIA_{\rm OS}$} \\
\hline
CE   &  -0.3$\pm$0.3    &   0.94$\pm$0.01 & -0.66\\
HST  &  2.4$\pm$2.3     &   1.08 $\pm$0.08 & -0.12\\
\hline
\multicolumn{4}{c}{$HIA_{\rm ISOS}$} \\
\hline
CE   &  -0.5$\pm$0.5   &   0.92$\pm$0.02 & -0.60\\
HST  &  -0.2$\pm$1.4    &   0.99$\pm$0.05 & -0.17\\
 \hline
\end{tabular}
\tablefoot{The regressions were conducted by the SciPy package.\\
\tablefoottext{$\dag$} {The correlation coefficient assumes values within a range from -1 (indicating a perfect negative correlation) to +1 (indicating a perfect positive correlation). A correlation coefficient of zero denotes no relationship between the two variables under consideration.}}
\end{table}

\section{Summary}

We conducted deep observations of  the ammonia hyperfine intensity anomalies (HIAs) with the Effelsberg 100\,m telescope in fifteen infall source candidates.
By adopting a rational definition of the HIA proposed by \citet{2020A&A...640A.114Z}, we seek to test whether HIAs can be used to trace infall motions, in particular of the cold molecular gas.
Due to long integration times on the targets, all the NH$_{3}$ satellite lines are clearly detected and all the HIAs of the inner ($HIA_{\rm IS}$) and outer ($HIA_{\rm OS}$) satellite lines are derived.
In 14 out of 15 targets, either $HIA_{\rm IS}$ or $HIA_{\rm OS}$ values deviate from unity (indicating an anomaly) by more than their 1-$\sigma$ uncertainties. In 10 targets, both $HIA_{\rm IS}$ and $HIA_{\rm OS}$ values exceed their 1-$\sigma$ uncertainties. Thus the presence of HIAs is prevalent in our sample.
Meanwhile, all the derived HIAs remain within the framework of the existing two models, the hyperfine selective trapping (HST) and  systematic contraction or expansion motions (CE) models.
It is found that a majority of the HIAs in the sources likely harboring infall motions are still consistent with the HST model.
In three sources, HIAs are consistent with infall motions under the CE model, while a procedure mitigating effects of the HST model even uncovers six such sources.
Therefore, HIAs could be used as an infall tracer but seem to be not highly sensitive to infall motions in our single-dish observations.
Nevertheless, akin to the case of blue-skewed profiles, HIAs might be blended with emission from outflow activities, since ammonia emission is also commonly seen in outflows.

HIAs induced by the HST model are expected to be enhanced with increasing  kinetic temperatures ($T_{\rm K}$).
$HIA_{\rm IS}$ and $HIA_{\rm OS}$ in our observations may show higher anomalies with increasing $T_{\rm K}$, but the correlations are weak.
On the contrary, HIAs induced by infall motions seem to show relatively constant values against $T_{\rm K}$, suggesting that HIAs might serve as  more effective infall tracers for relatively cold gas.
High-resolution observations of HIAs are crucial to further constrain the origin of HIAs and assess the contributions from infall and outflow motions.

\begin{acknowledgements}
We thank the anonymous referee for useful suggestions improving the paper. Based on observations with the 100-m telescope of the MPIfR (Max-Planck-Institut f\"{u}r Radioastronomie) at Effelsberg. This work was funded by the National Key R\&D Program of China (No. 2022YFA1603103), the CAS ``Light of West China'' Program (No. 2021-XBQNXZ-028), the National Natural Science foundation of China (Nos.12103082, 11603063, and 12173075), and the Natural Science Foundation of Xinjiang Uygur Autonomous Region (No. 2022D01A362).
GW acknowledges the support from Youth Innovation Promotion Association CAS.
\end{acknowledgements}

%   \bibliography{ref.bib} % your references Yourfile.bib

\begin{appendix}
\onecolumn

\section{Observed NH$_{3}$ $(J,K) = (1,1)$ and $(2,2)$ spectra.}
\label{app-specs}

    \begin{figure*}[h]
    \centering
    \includegraphics[width=0.33\hsize]{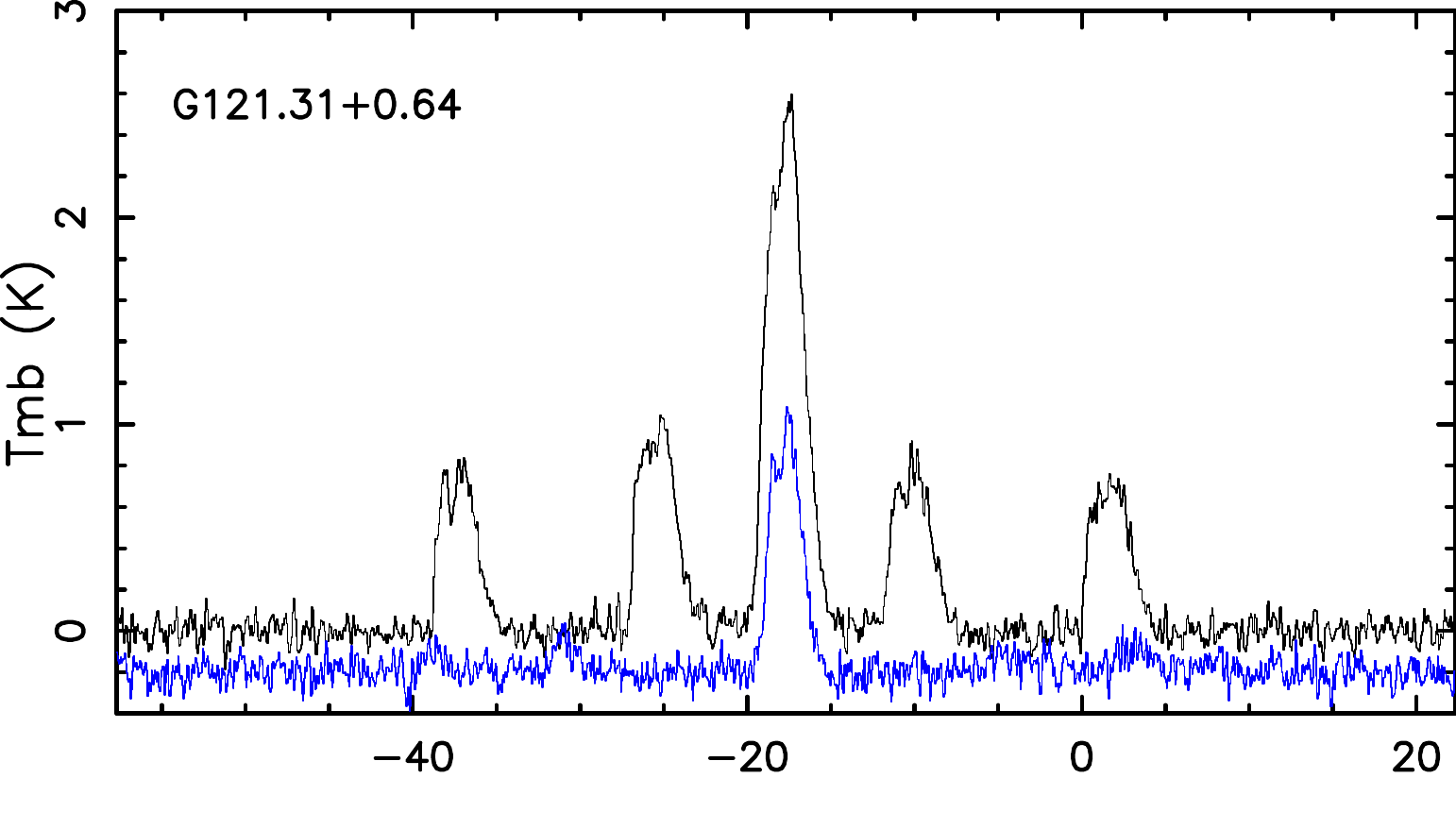}
    \includegraphics[width=0.33\hsize]{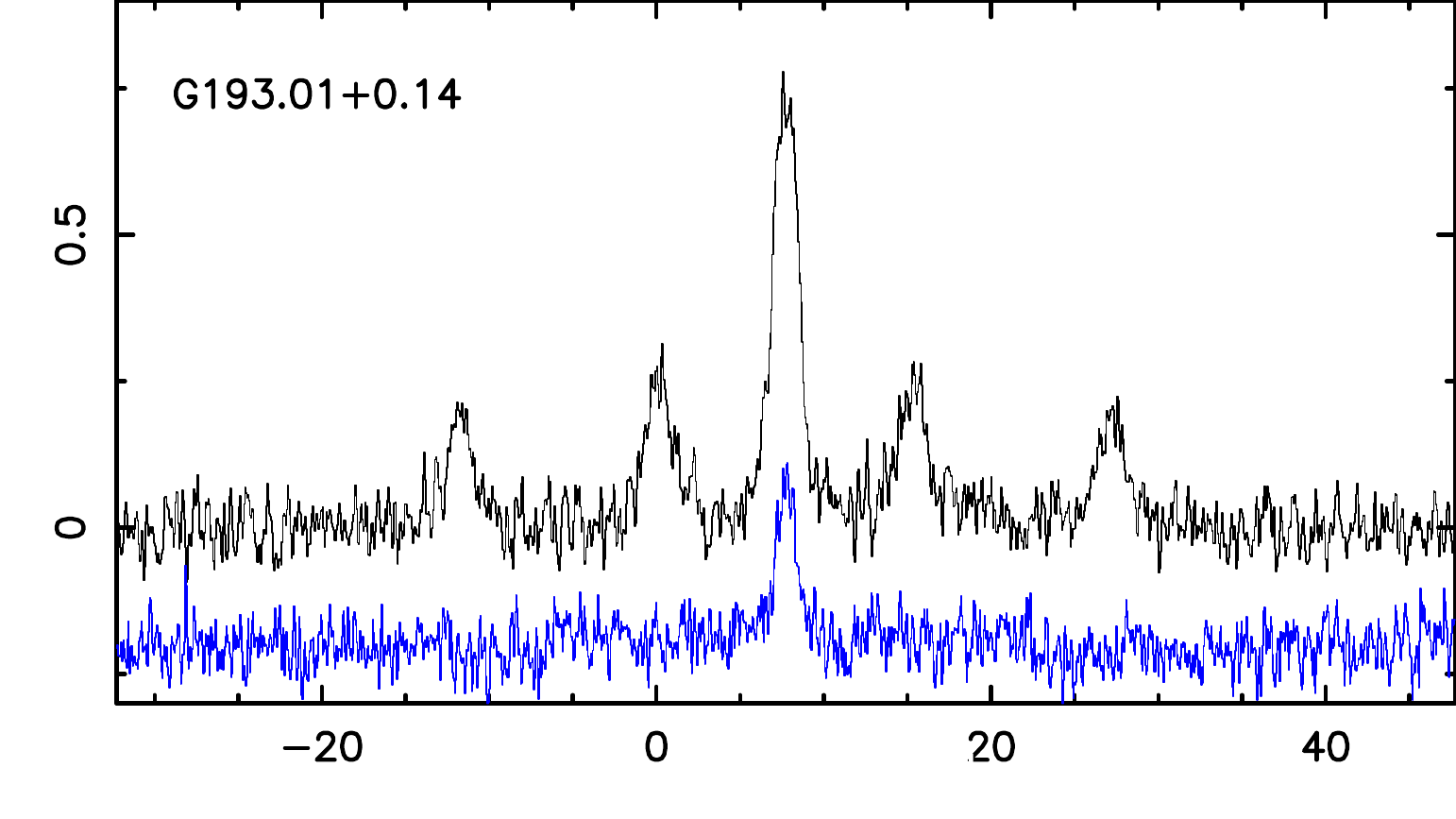}
    \includegraphics[width=0.33\hsize]{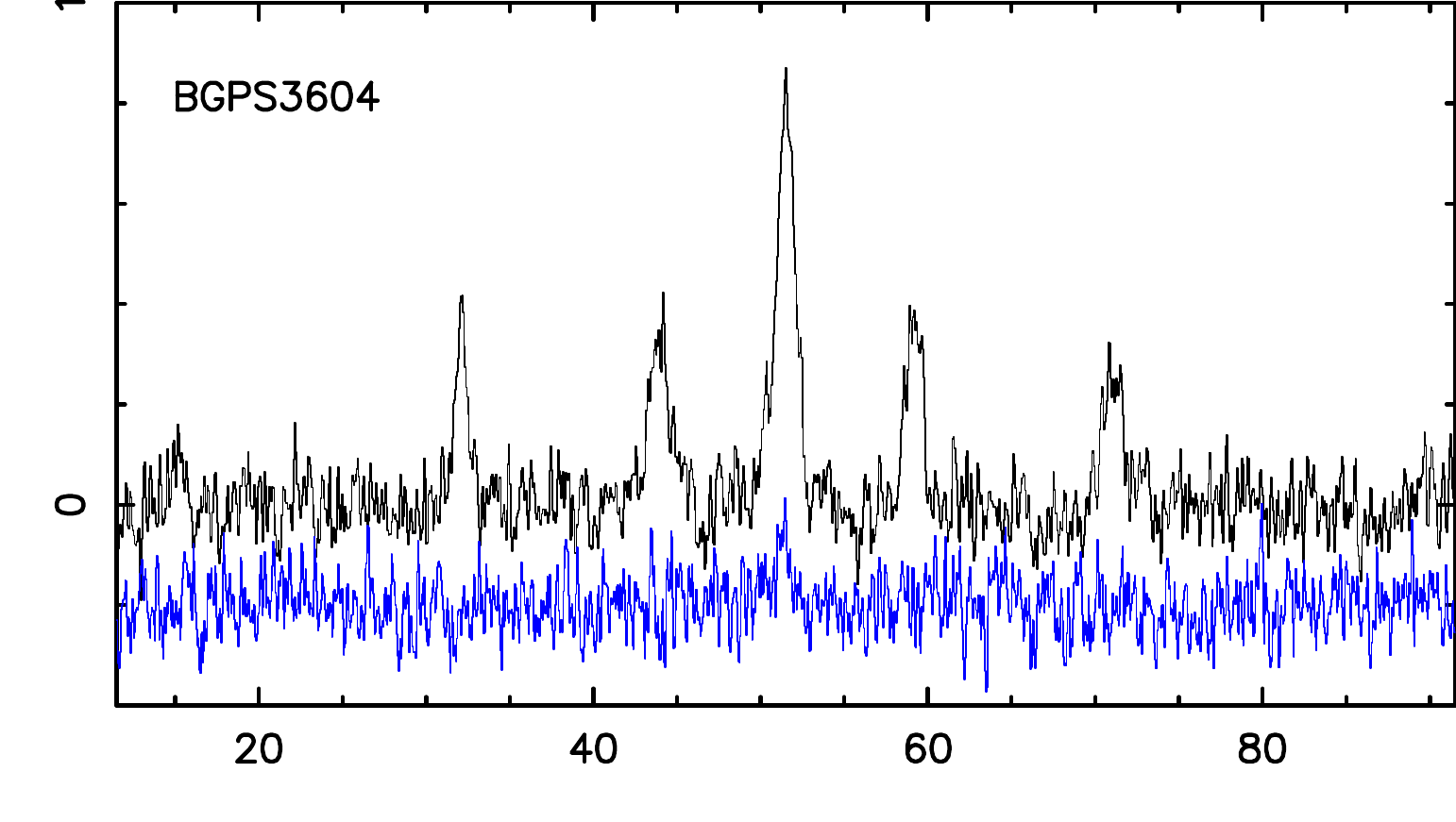}
    \includegraphics[width=0.33\hsize]{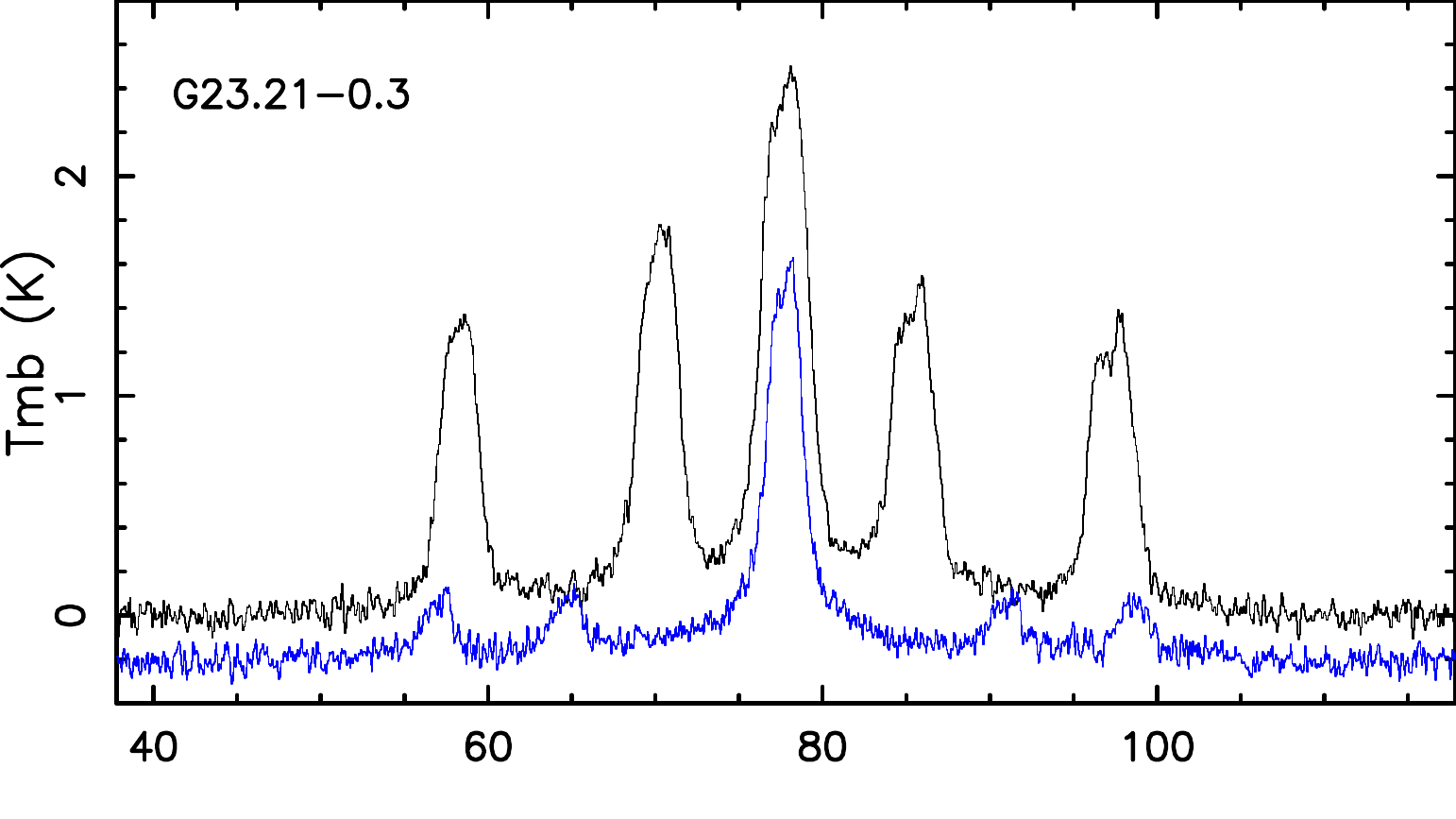}
    \includegraphics[width=0.33\hsize]{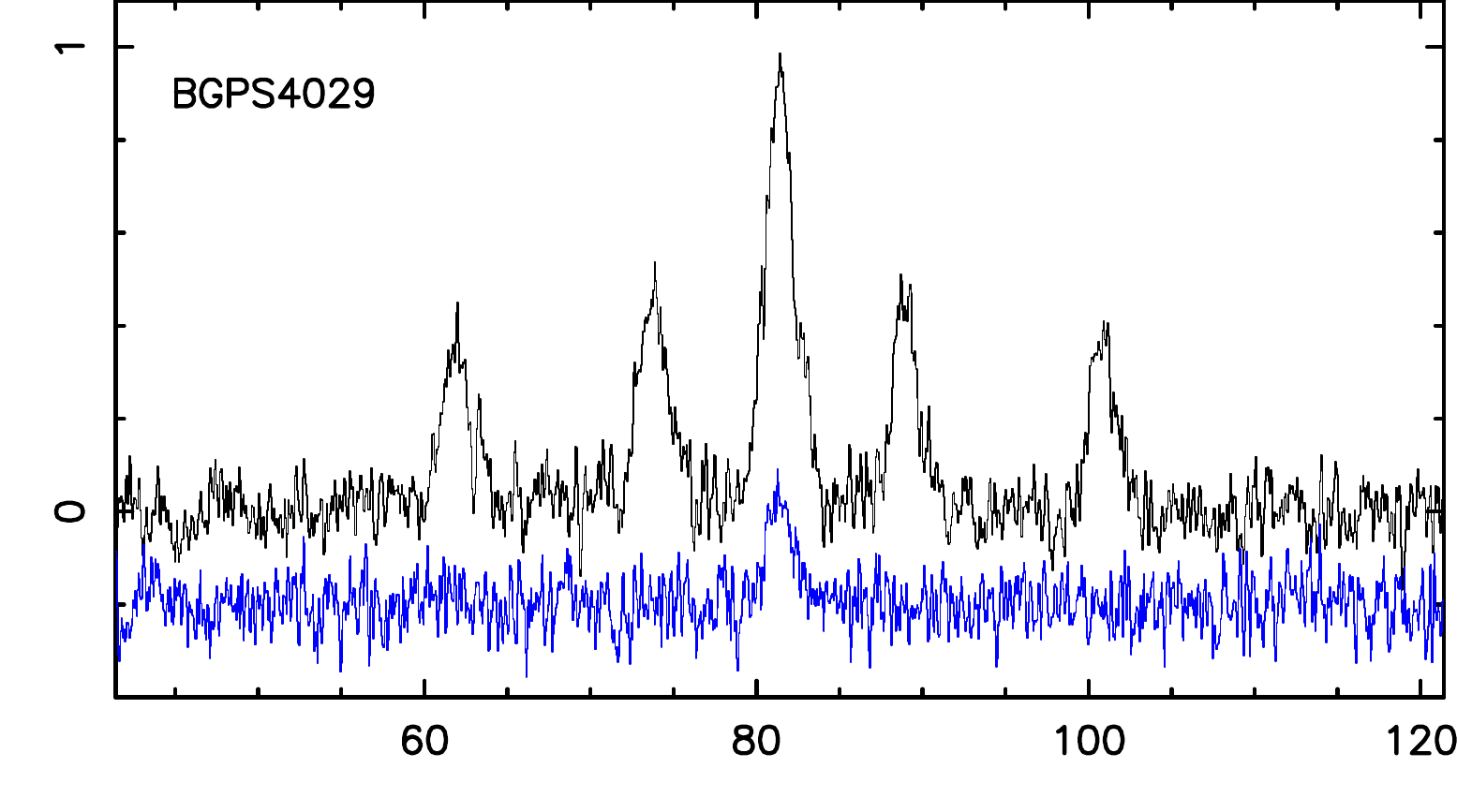}
    \includegraphics[width=0.33\hsize]{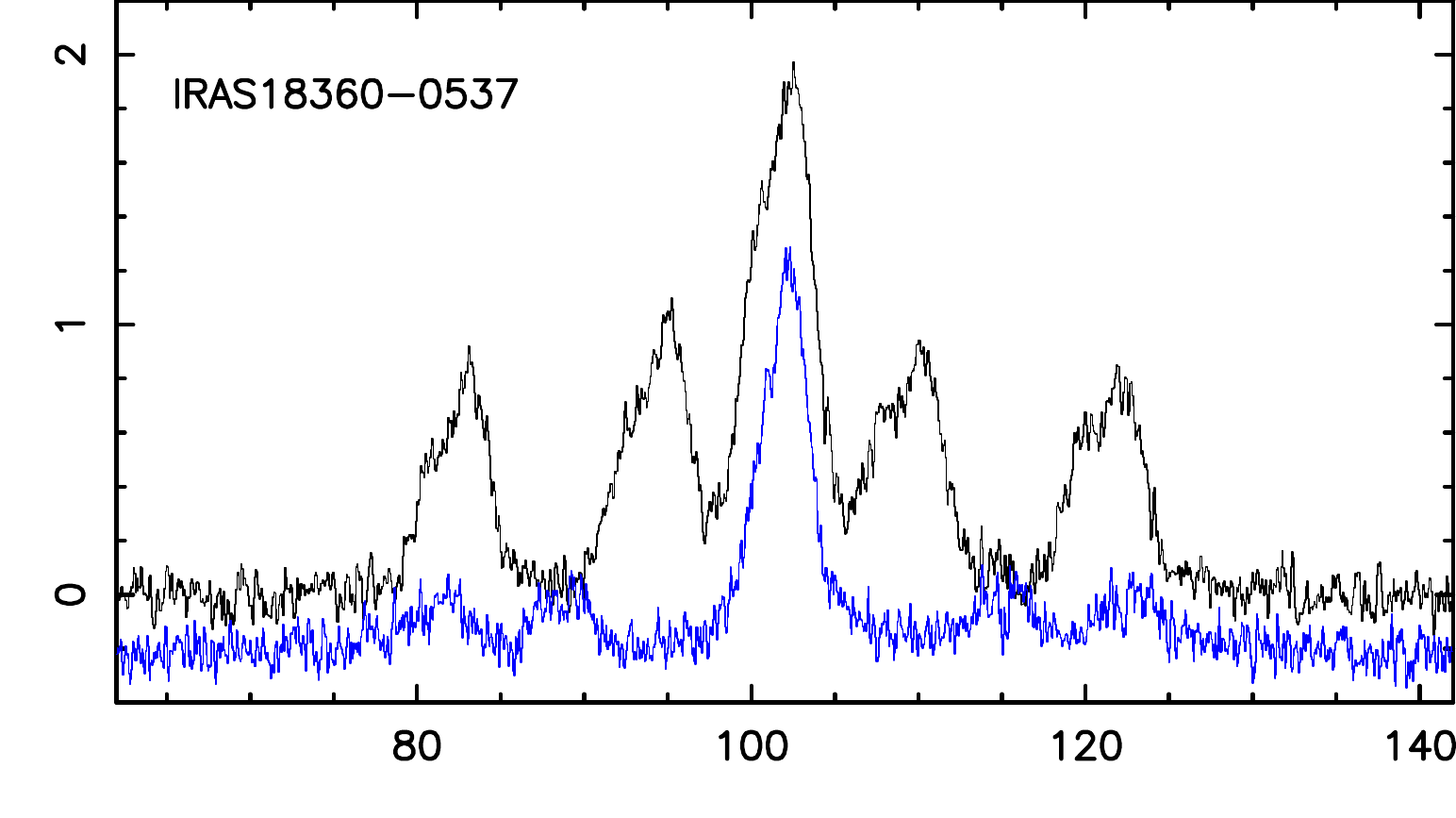}
    \includegraphics[width=0.33\hsize]{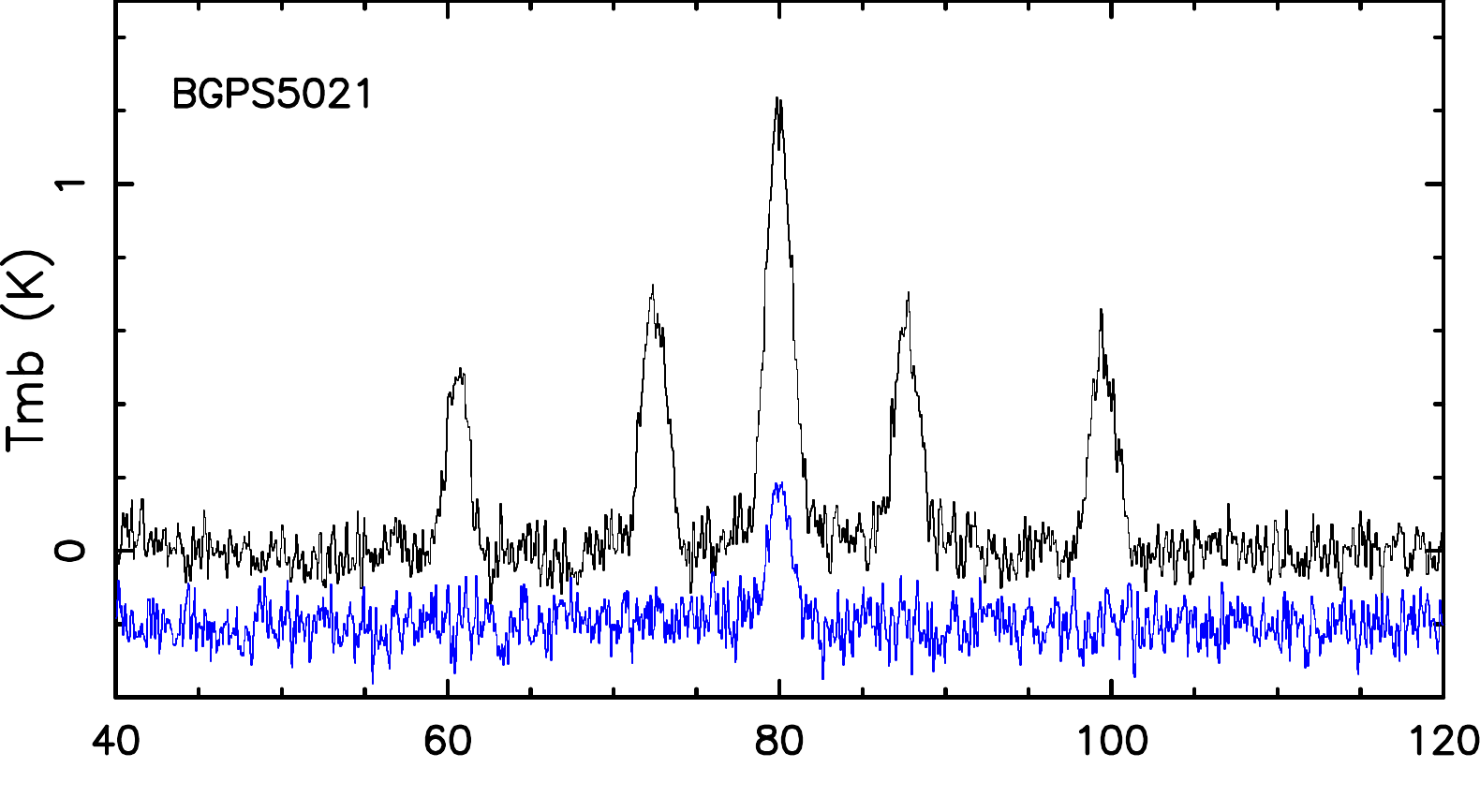}
    \includegraphics[width=0.33\hsize]{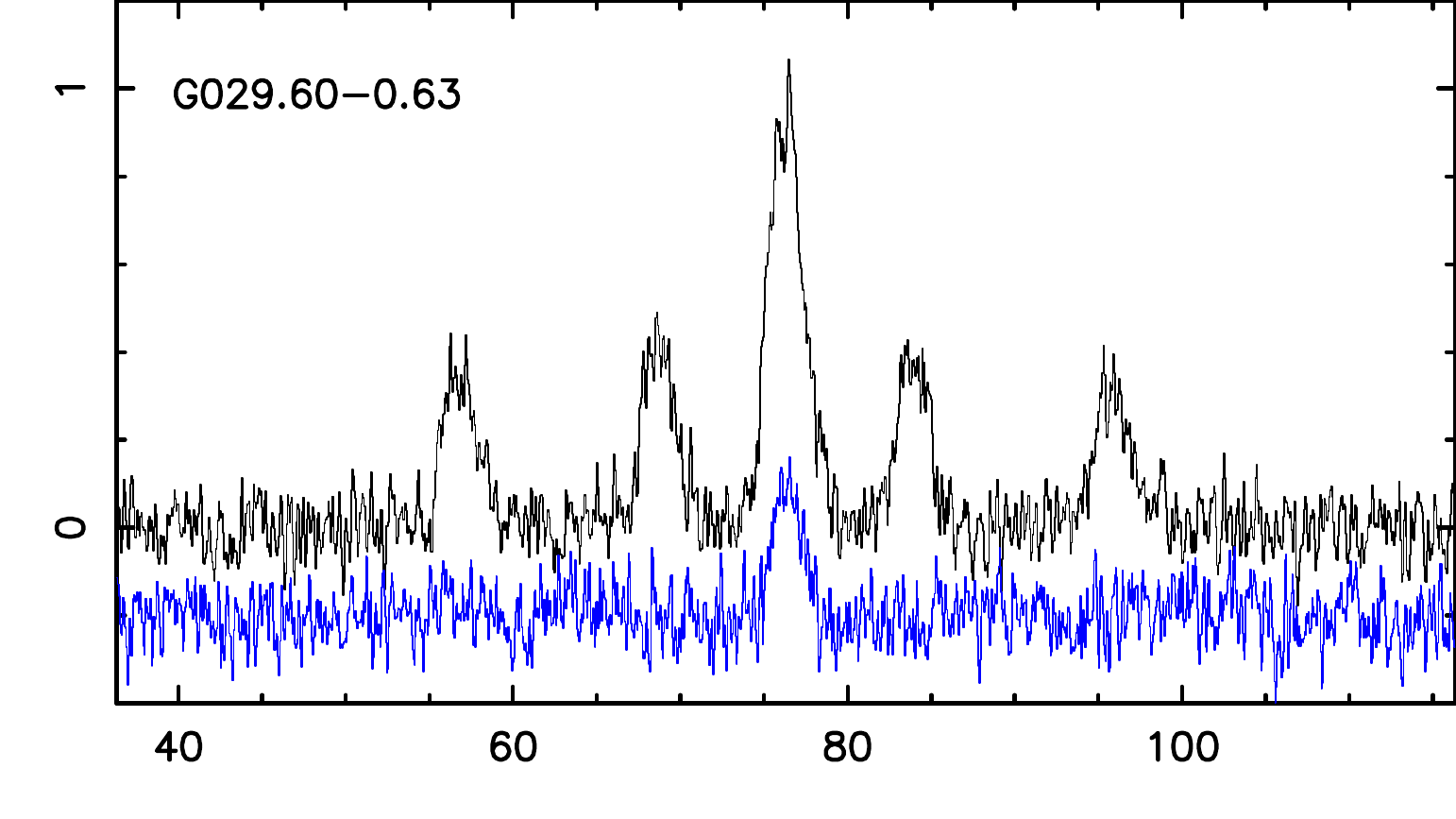}
    \includegraphics[width=0.33\hsize]{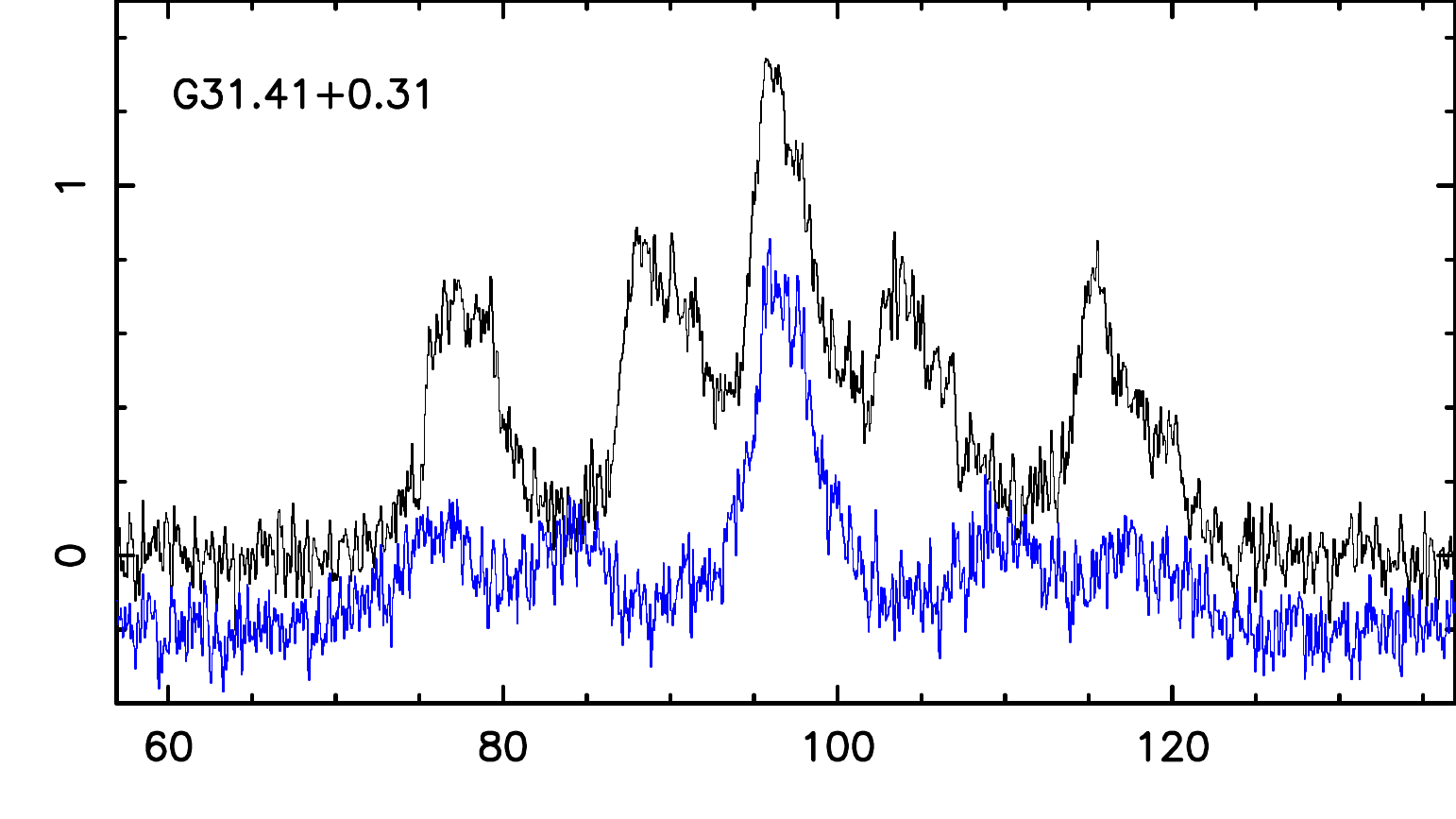}
    \includegraphics[width=0.33\hsize]{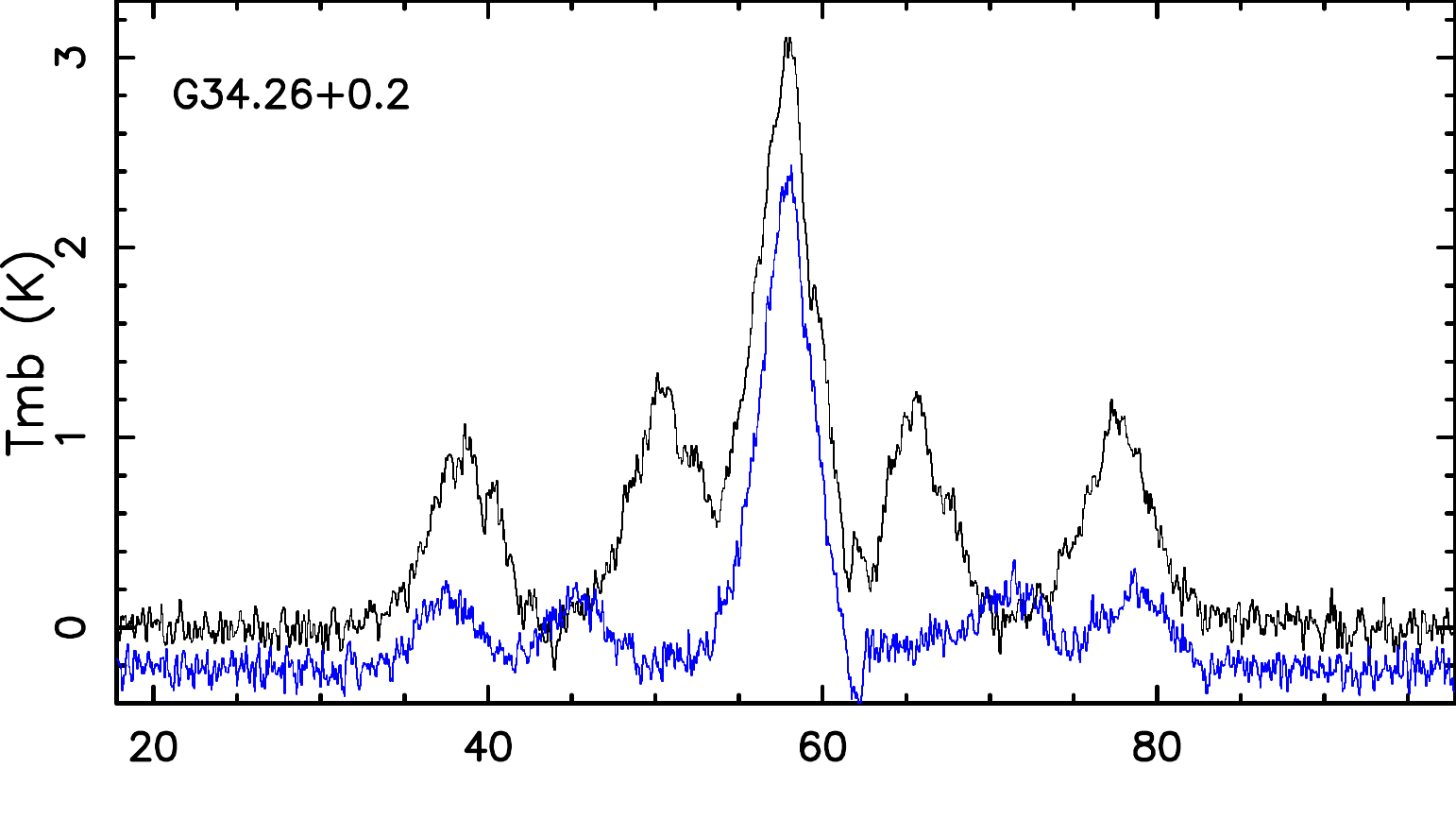}
    \includegraphics[width=0.33\hsize]{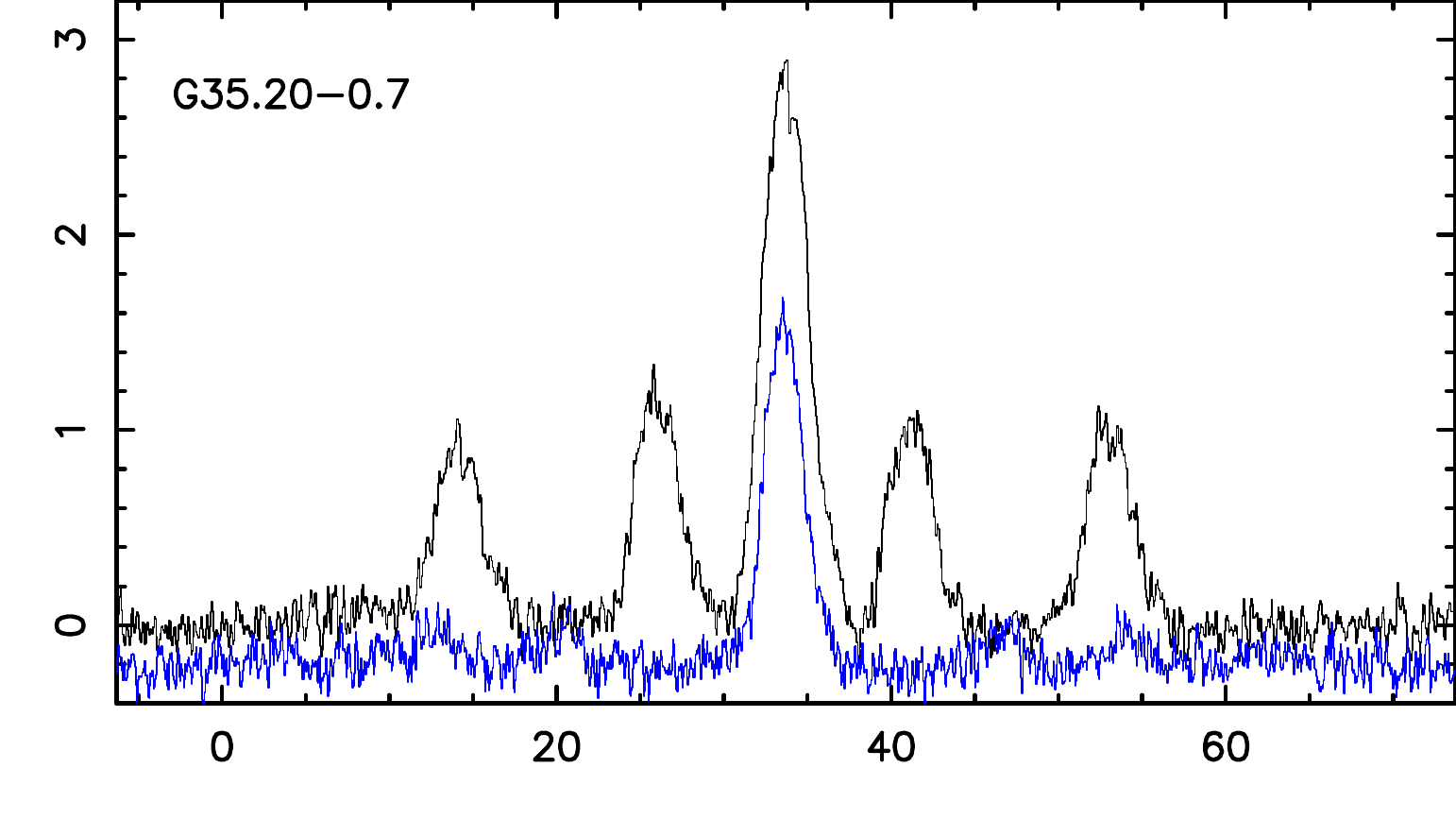}
    \includegraphics[width=0.33\hsize]{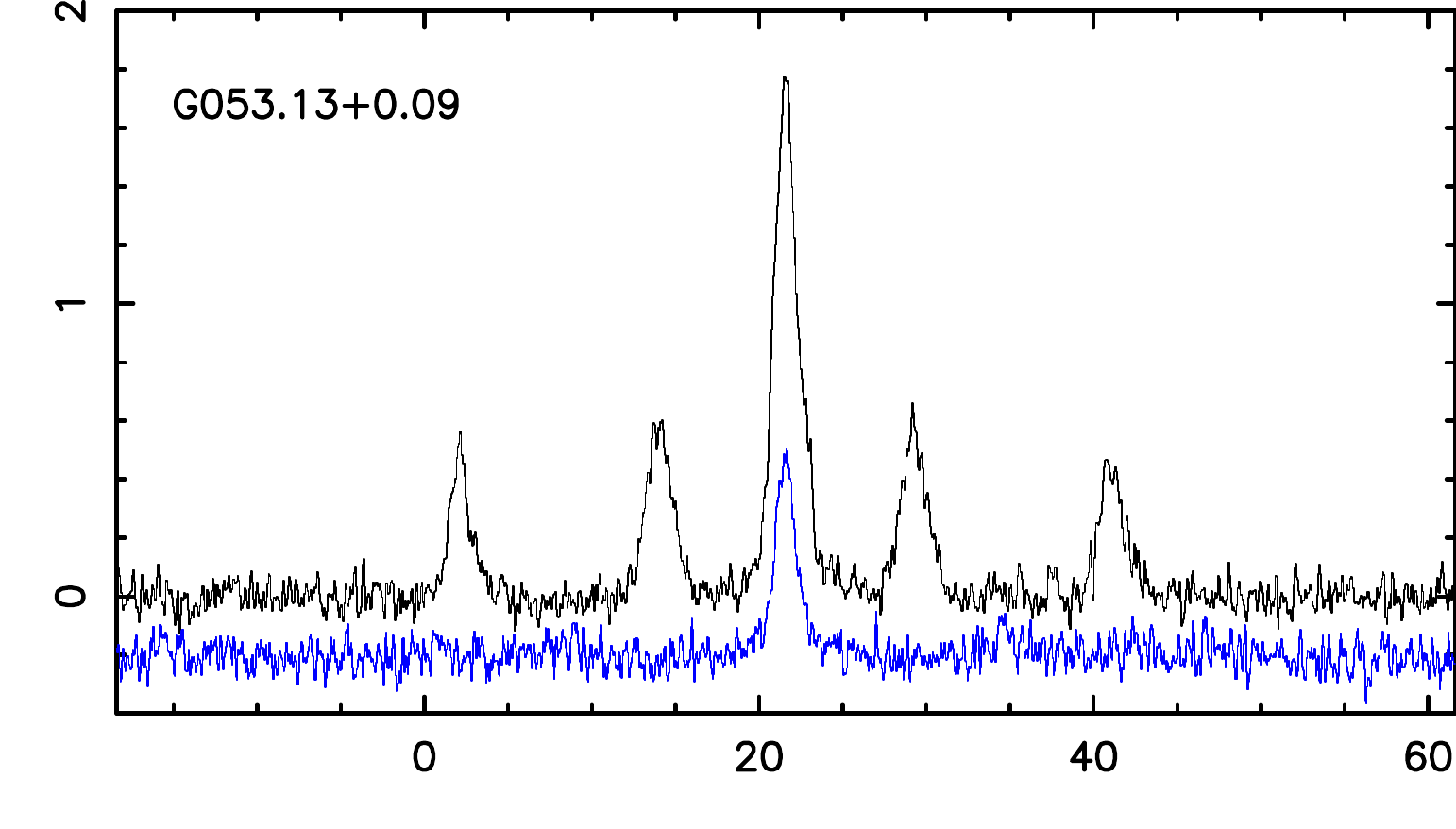}
    \includegraphics[width=0.33\hsize]{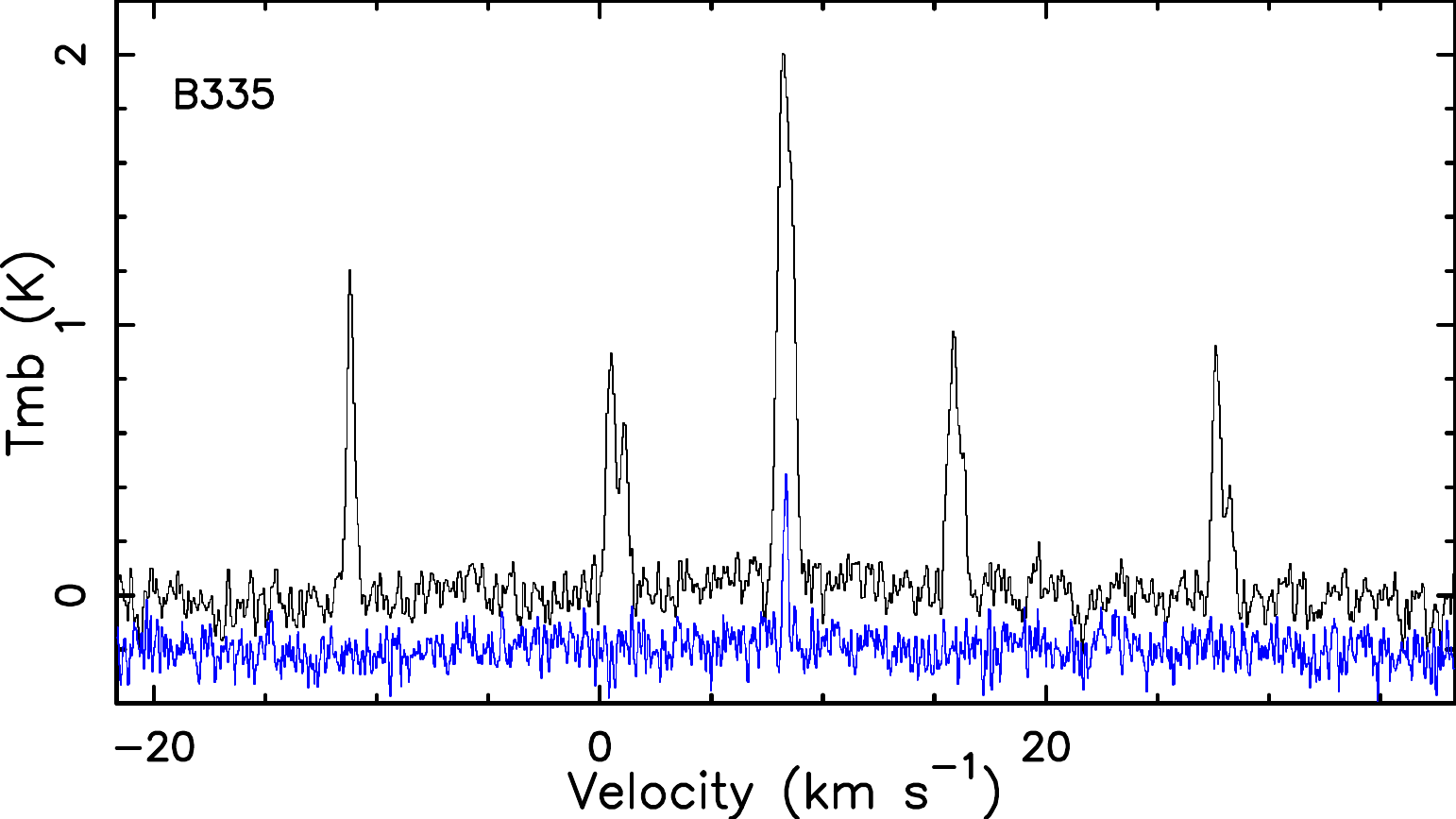}
    \includegraphics[width=0.33\hsize]{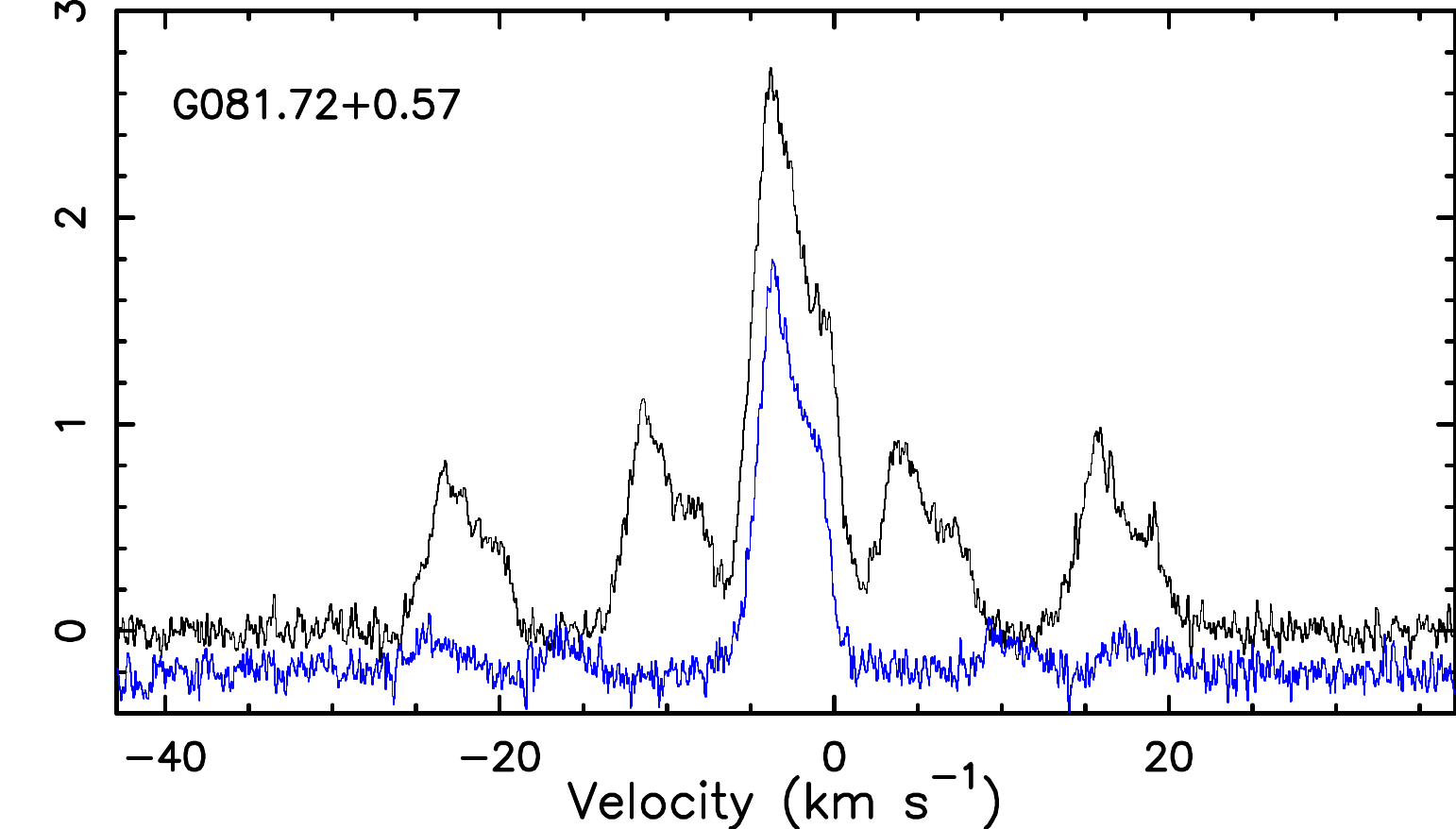}
    \includegraphics[width=0.33\hsize]{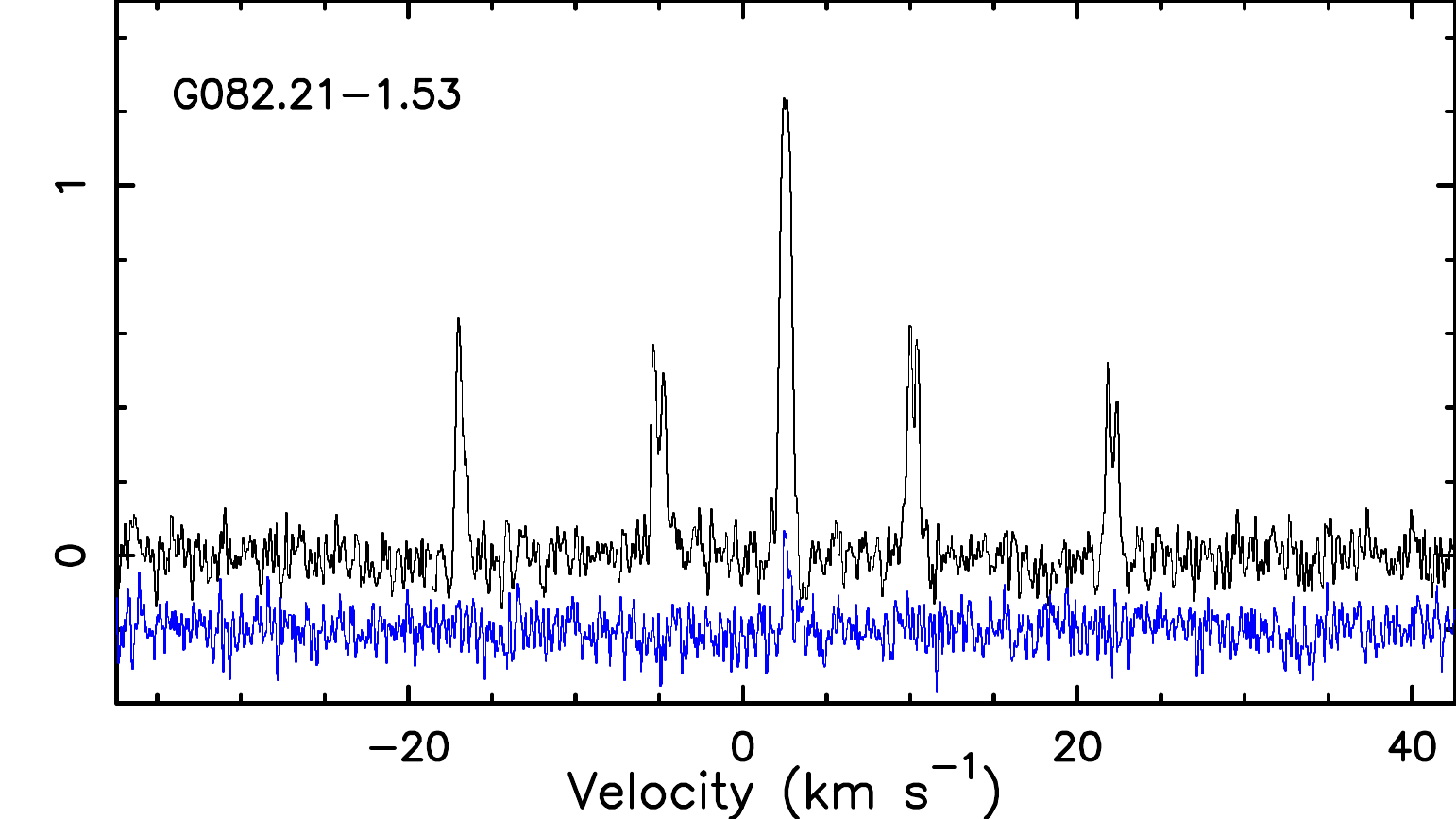}
    \caption{NH$_{3}$ $(J,K) = (1,1)$ and $(2,2)$ spectra. The observed NH$_{3}$ $(J,K) = (1,1)$ and $(2,2)$ spectra are shown in black and blue colors in each panel. NH$_{3}$ $(J,K) = (2,2)$ spectra are shifted to -0.2 K on the Y-axis. The source name is labeled in the top-left corner of each panel.}
    \label{fig:spec112233}
    \end{figure*}

\end{appendix}

\end{document}